\documentclass[12pt]{article}

\usepackage{epsfig,cite}

\catcode`@=11

\begin{document}



\input paperdef 

\graphicspath{{figssmall/}}


\thispagestyle{empty}
\setcounter{page}{0}
\def\thefootnote{\fnsymbol{footnote}}

\begin{flushright}
DCPT/07/12\\
IPPP/07/06\\
arXiv:0704.0619 \\
\end{flushright}

\mbox{}\vspace{0em}

\begin{center}

{\large\sc {\bf Search for Heavy Neutral MSSM Higgs Bosons with CMS:}}

\vspace*{0.3cm}

{\large\sc {\bf Reach and Higgs-Mass Precision}}

\vspace{0.5cm}

{\sc S.~Gennai$^{\,1}$%
\footnote{
email: Simone.Gennai@cern.ch
}%
, S.~Heinemeyer$^{\,2}$%
\footnote{
email: Sven.Heinemeyer@cern.ch
}%
, A.~Kalinowski$^{\,3}$%
\footnote{
email: Artur.Kalinowski@cern.ch
}%
, R.~Kinnunen$^{\,4}$%
\footnote{
email: Ritva.Kinnunen@cern.ch
}%
, S.~Lehti$^{\,4}$%
\footnote{
email: Sami.Lehti@cern.ch
}%
, \\[.5em] 
A.~Nikitenko$^{\,5}$%
\footnote{
email: Alexandre.Nikitenko@cern.ch
}%
~and G.~Weiglein$^{\,6}$%
\footnote{
email: Georg.Weiglein@durham.ac.uk
}%
}

\vspace*{0.5cm}

$^1$ Centro Studi Enrico Fermi, Rome and INFN Pisa, Italy

\vspace*{0.1cm}

$^2$ Instituto de Fisica de Cantabria (CSIC-UC), Santander, Spain

\vspace*{0.1cm}

$^3$ Institute of Experimental Physics, Warsaw, Poland

\vspace*{0.1cm}

$^4$ Helsinki Institute of Physics, Helsinki, Finland

\vspace*{0.1cm}

$^5$ Imperial College, London, UK; on leave from ITEP, Moscow, Russia

\vspace*{0.1cm}

$^6$ IPPP, University of Durham, Durham DH1~3LE, UK

\end{center}

\vspace*{0.1cm}

\begin{abstract}
The search for MSSM Higgs bosons will be an important goal at the
LHC. We analyze the search reach of the CMS experiment for the heavy
neutral MSSM Higgs bosons with an integrated luminosity of 30 or
60~\ifb. This is done by combining the latest results for the CMS
experimental sensitivities based on full simulation studies with
state-of-the-art theoretical predictions of MSSM Higgs-boson properties.
The results are interpreted in MSSM benchmark scenarios in terms of the 
parameters $\tb$ and the Higgs-boson mass scale, $\MA$. We study the
dependence of the 5~$\si$ discovery contours in the $\MA$--$\tb$ plane
on variations of the other supersymmetric parameters. 
The largest effects arise from a change in the higgsino mass parameter $\mu$, 
which enters both via higher-order radiative corrections and via the 
kinematics of Higgs decays into supersymmetric particles. While the 
variation of $\mu$ can shift the prospective discovery reach (and
correspondingly the ``LHC wedge'' region) by about $\De\tb = 10$, we
find that the discovery reach is rather stable with respect to the impact of
other supersymmetric parameters.
Within the discovery region we analyze the accuracy with which the 
masses of the heavy neutral Higgs bosons can be determined. We find that
an accuracy of 1--4\% should be achievable, which could make it
possible in favourable regions of the MSSM parameter space to 
experimentally resolve the signals of the two heavy MSSM Higgs bosons at
the LHC.
\end{abstract}

\def\thefootnote{\arabic{footnote}}
\setcounter{footnote}{0}

\newpage


\section{Introduction}

Identifying the mechanism of electroweak symmetry
breaking will be one of the main goals of the LHC. 
Many possibilities have been studied in the literature, of which 
the most popular ones are the Higgs mechanism within the Standard Model (SM) 
and within the Minimal Supersymmetric Standard Model
(MSSM)~\cite{mssm}. Contrary to the case of the SM, in the MSSM 
two Higgs doublets are required.
This results in five physical Higgs bosons instead of the single Higgs
boson of the SM. These are the light and heavy $\cp$-even Higgs bosons, $h$
and $H$, the $\cp$-odd Higgs boson, $A$, and the charged Higgs boson,
$H^\pm$.%
\footnote{We focus in this paper on the case without explicit $\cp$-violation 
in the soft supersymmetry-breaking terms.}
The Higgs sector of the MSSM can be specified at lowest
order in terms of the gauge couplings, the ratio of the two Higgs vacuum
expectation values, $\tb \equiv v_2/v_1$, and the mass of the $\cp$-odd
Higgs boson, $\MA$.
Consequently, the masses of the $\cp$-even neutral Higgs bosons and the
charged Higgs boson are dependent quantities that can be
predicted in terms of the Higgs-sector parameters. Higgs-phenomenology
in the MSSM is strongly affected by higher-order corrections, in
particular from the sector of the third generation quarks and squarks,
so that the dependencies on various other MSSM parameters can be
important.

After the termination of LEP in the year 2000 (the final LEP
results can be found in \citeres{LEPHiggsSM,LEPHiggsMSSM}), and the
(ongoing) Higgs boson search at the 
Tevatron~\cite{D0bounds,CDFbounds,Tevcharged}, the search
will be continued at the LHC~\cite{lhctdrsA,atlashiggs,lhctdrsS}
(see also \citeres{jakobs,schumi} for recent reviews). The current
exclusion bounds within the MSSM~\cite{LEPHiggsMSSM,D0bounds,CDFbounds}
and the prospective sensitivities at
the LHC are usually displayed in terms of the parameters $\MA$ and $\tb$
that characterize the MSSM Higgs sector at lowest order. The other MSSM
parameters are conventionally fixed according to certain benchmark
scenarios~\cite{benchmark,benchmark2,benchmark3}. 
The most prominent one is the ``$\mhmax$~scenario'', which 
in the search for the light $\cp$-even Higgs boson
allows to obtain conservative bounds on $\tb$ for fixed values of the
top-quark mass and the scale of the supersymmetric particles~\cite{tbexcl}. 
Besides the ``no-mixing scenario'', which is similar to the
$\mhmax$~scenario, but assumes vanishing mixing in the stop sector,
other $\cp$-conserving scenarios that have been studied in LHC
analyses (see e.g.\ \citere{schumi}) are the ``gluophobic Higgs scenario'' 
and the ``small~$\aeff$''~scenario~\cite{benchmark2}.

For the interpretation of the exclusion bounds and prospective discovery
contours in the benchmark scenarios it is important to assess how
sensitively the results depend on those parameters that have been fixed
according to the benchmark prescriptions.
While in the decoupling limit, which is the region of MSSM parameter
space with $\MA \gg \MZ$, the couplings of the light $\cp$-even Higgs boson 
approach those of a SM Higgs boson with the same mass, the couplings of
the heavy Higgs bosons of the MSSM can be sizably affected by
higher-order contributions even for large values of $\MA$. The
kinematics of the heavy Higgs-boson production processes, on the other hand, 
is governed by the parameter $\MA$, since in the region of large $\MA$
the heavy MSSM Higgs bosons are nearly mass-degenerate, 
$\MA \approx \MH \approx \MHp$. In \citere{benchmark3} it has been
shown that higher-order contributions to the relation between the
bottom-quark mass and the bottom-Yukawa coupling have a dramatic effect
on the exclusion bounds in the $\MA$--$\tb$ plane obtained from the 
$b \bar b \phi, \phi \to b \bar b$ channel at the Tevatron.

In this article we investigate how the 
5$\,\si$ discovery regions in the $\MA$--$\tb$ plane 
for the heavy neutral MSSM Higgs bosons
(a corresponding analysis for the charged Higgs-boson search will be 
presented elsewhere)
obtainable with the CMS experiment at the LHC depend on the other MSSM
parameters.
For the experimental sensitivities
achievable with CMS we use up-to-date results based on full simulation
studies for 30 or 60~\ifb (depending on the channel)~\cite{lhctdrsS}.
This information is combined with precise theory
predictions for the Higgs-boson masses and the involved production and
decay processes incorporating higher-order
corrections at the one-loop and two-loop level. In our analysis we
investigate the impact on the discovery reach arising both from 
higher-order corrections and from possible decays of the heavy Higgs bosons
into supersymmetric particles.%
\footnote{We restrict our analysis to the impact of 
supersymmetric contributions. For a discussion of 
uncertainties related to parton distribution functions, see
e.g.\ \citere{Belyaev:2005nu}.}

The search for the heavy neutral MSSM Higgs bosons at the LHC will mainly be
pursued in the $b$~quark associated production with a subsequent decay to
$\tau$~leptons~\cite{lhctdrsA,atlashiggs,lhctdrsS}. In the region of
large $\tb$ this production process benefits from an enhancement factor 
of $\TQb$ compared to the SM case. 
The main search channels are%
\footnote{In our analysis we do not consider diffractive
Higgs production, $pp \to p \oplus H \oplus p$~\cite{diffHSM}.
For a detailed discussion of the search reach for the heavy neutral MSSM 
Higgs bosons in diffractive Higgs production we refer to
\citere{diffHMSSM}.}
(here and in the
following $\phi$ denotes the two heavy neutral MSSM Higgs bosons, 
$\phi = H, A$):
\BEA
\label{jj}
    && b \bar b \phi,  \phi \to \tau^+\tau^- \to 2 \, \mbox{jets}\\[.3em]
\label{mj}
    && b \bar b \phi,  \phi \to \tau^+\tau^- \to \mu + \, \mbox{jet}\\[.3em]
\label{ej}
    && b \bar b \phi,  \phi \to \tau^+\tau^- \to e + \,\mbox{jet}\\[.3em]
\label{em}
    && b \bar b \phi,  \phi \to \tau^+\tau^- \to e + \mu ~.
\EEA

For our numerical analysis we use the program
{\tt FeynHiggs}~\cite{feynhiggs,mhiggslong,mhiggsAEC,feynhiggs2.5}.
We study in particular the dependence of the ``LHC wedge'' region, i.e.\
the region in which only the light $\cp$-even  MSSM Higgs boson can be
detected at the LHC at the 5$\,\si$ level,
on the variation of the higgsino mass parameter $\mu$. The dependence on
$\mu$ enters in two different ways, on the one hand via higher-order
corrections affecting the relation between the bottom mass and the
bottom Yukawa coupling, and on the other hand
via the kinematics of Higgs decays into supersymmetric particles. 
We analyze both effects separately and discuss the possible impact of
other supersymmetric parameters.

Our results for the discovery reach of the heavy neutral MSSM Higgs
bosons extend the known results in the literature in various ways.
In comparison with \citeres{cmshiggs,KinNik}, where the 
prospective $5 \si$ discovery contours
for CMS in the $\MA$--$\tb$ plane of the $\mhmax$ benchmark scenario
were given for three different values of $\mu$, 
the results in the present paper 
are based on full simulation
studies and make use of the most up-to-date CMS tools for triggering and
event reconstruction. Furthermore, in the analysis of
\citeres{cmshiggs,KinNik} relevant higher-order corrections, in
particular those depending on $\db$ (see \refse{sec:higherorder} below), 
have been neglected.
The effects induced by the $\db$ corrections have been investigated in
\citere{benchmark3}, where the results were obtained 
by a simple rescaling of the experimental results given in
\citeres{lhctdrsA,cmshiggs,KinNik,atlasnote}. Our present analysis, on
the other hand, makes use of the latest CMS studies and provides a
separate treatment of the different $\tau$ final states, channels 
(\ref{jj})--(\ref{em}). 

As a second step of our analysis we investigate the 
experimental precision that can be achieved for the
determination of the heavy Higgs-boson masses in the 
discovery channels (\ref{jj})--(\ref{em}). We discuss the prospective 
accuracy of the mass measurement in view of the possibility to
experimentally resolve the signals of the heavy neutral MSSM Higgs 
bosons.

The paper is organized as follows: \refse{sec:theory} introduces our
notation and gives a brief summary of the most relevant supersymmetric
radiative corrections to the Higgs-boson masses, production cross
sections and decay widths at the LHC.
The relevant benchmark scenarios are briefly reviewed.
In \refse{sec:expanal} the experimental analysis is described.
The results for the variation of the 5$\,\si$ discovery contours, 
obtainable at CMS with
30 or 60~\ifb\ are given in \refse{sec:results}, where we also discuss the
achievable experimental precision in the Higgs mass determination. 
The conclusions can be found in \refse{sec:conclusions}.


\section{Phenomenology of the MSSM Higgs sector}
\label{sec:theory}

\subsection{Notation}

The MSSM Higgs sector at lowest order is described in terms of two  
independent parameters (besides the SM gauge couplings): 
$\tb \equiv v_2/v_1$, the ratio of the two vacuum expectation values, and 
$M_A$, the mass of the $\cp$-odd Higgs boson~$A$.
Beyond the tree-level, large radiative corrections can occur from the
$t/\Stop$ sector, and for large values of $\tb$ also from the 
$b/\Sbot$ sector.

Our notations for the scalar top and scalar bottom sector of the MSSM
are as follows:
the mass matrices in the basis of the current eigenstates $\StopL, \StopR$ and
$\SbotL, \SbotR$ are given by
\BEA
\label{stopmassmatrix}
{\cal M}^2_{\Stop} &=&
  \ML \MSQ^2 + \mt^2 + \CZb (\edz - \frac{2}{3} \sw^2) \MZ^2 &
      \mt \Xt \\
      \mt \Xt &
      \MstR^2 + \mt^2 + \frac{2}{3} \CZb \sw^2 \MZ^2 
  \MR, \\
&& \non \\
\label{sbotmassmatrix}
{\cal M}^2_{\Sbot} &=&
  \ML \MSQ^2 + \mb^2 + \CZb (-\edz + \frac{1}{3} \sw^2) \MZ^2 &
      \mb \Xb \\
      \mb \Xb &
      \MsbR^2 + \mb^2 - \frac{1}{3} \CZb \sw^2 \MZ^2 
  \MR,
\EEA
where 
\BE
\mt \Xt = \mt (\At - \mu \CTb) , \quad
\mb\, \Xb = \mb\, (\Ab - \mu \Tb) .
\label{eq:mtlr}
\EE
Here $\MSQ$, $\MstR$ and $\MsbR$ are the diagonal soft SUSY-breaking
parameters, $\At$ denotes the trilinear Higgs--stop coupling, $\Ab$
denotes the 
Higgs--sbottom coupling, and $\mu$ is the higgsino mass parameter.

For the numerical evaluation, it is often convenient to choose
\BE
\MSQ = \MstR = \MsbR =: \msusy .
\label{eq:msusy}
\EE
Concerning analyses for the case where $\MstR \neq \MSQ \neq \MsbR$, see
e.g.\ \citeres{stefanCM,mhiggslong,mhiggsEP3b}. It has been shown that the
upper bound on the mass of the light $\cp$-even Higgs boson, $\Mh$, 
obtained using \refeq{eq:msusy} is the same as for the more general
case, provided that $\msusy$ is identified with the heaviest mass of
$\MSQ, \MstR, \MsbR$~\cite{mhiggslong}. 

Accordingly, the most important parameters entering the Higgs-sector
predictions via higher-order corrections are
$\mt$, $\msusy$, $\Xt$, $\Xb$ and $\mu$ (see also the discussion in
\refse{sec:db} below). The Higgs-sector observables
furthermore depend on the SU(2) gaugino mass parameter, $M_2$, the U(1)
parameter $M_1$ and the gluino mass, $\mgl$ (the latter enters the
predictions for the Higgs-boson masses only from two-loop order on).
In numerical analyses the U(1) gaugino mass parameter, $M_1$, is often
fixed via the GUT relation 
\BE
M_1 = \frac{5}{3} \frac{\sw^2}{\cw^2} M_2.
\EE
We will briefly comment below on the possible impact of complex phases
entering the Higgs-sector predictions via higher-order contributions.


\subsection{Higher-order corrections in the Higgs sector}
\label{sec:higherorder}

In the following we briefly summarize the most important higher-order 
corrections affecting the observables in the MSSM Higgs-boson sector. 
As mentioned above, we focus on the MSSM with real parameters. 
For our numerical analysis we use the program 
{\tt FeynHiggs}~\cite{feynhiggs,mhiggslong,mhiggsAEC,feynhiggs2.5}%
\footnote{
The code can be obtained from {\tt www.feynhiggs.de} .
}%
, which incorporates a comprehensive set of higher-order results 
obtained in the Feynman-diagrammatic 
approach~\cite{feynhiggs2.5,mhiggsletter,mhiggslong,mhiggsAEC,mhiggsEP4b,bse}.

\subsubsection{Higgs-boson propagator corrections}

Higher-order corrections to the Higgs-boson masses and the wave
function normalization factors of processes with external Higgs bosons 
arise from Higgs-boson propagator-type contributions. These corrections
furthermore contribute in a universal way to all Higgs-boson
couplings. For the propagator-type corrections in the MSSM the complete
one-loop results~\cite{ERZ,mhiggsf1lA,mhiggsf1lB,mhiggsf1lC}, the bulk
of the two-loop
contributions~\cite{mhiggsRefs,mhiggsletter,mhiggslong,mhiggsEP3b,deltamb1,deltamb2,deltamb2b,mhiggsEP4b,mhiggsEP5}
and even leading three-loop corrections~\cite{mhiggs3l} are known.
The remaining theoretical uncertainty on the light $\cp$-even Higgs-boson 
mass has been estimated to be below 
$\sim 3 \gev$~\cite{mhiggsAEC,PomssmRep}. 
The by far dominant contribution is the \order{\alt} term due to top and stop 
loops ($\alt \equiv h_t^2 / (4 \pi)$, where $h_t$ denotes the 
top-quark Yukawa coupling). Effects of \order{\alb} can be important
for large values of $\tb$. 


\subsubsection{Corrections to the relation between the bottom-quark mass
and the bottom Yukawa coupling}
\label{sec:db}

Concerning the corrections from the bottom/sbottom sector, large
higher-order effects can in particular occur in the relation between
the bottom-quark mass and the bottom Yukawa coupling (which controls the
interaction between the Higgs bosons and bottom quarks as well as
between the Higgs and scalar bottoms), $h_b$, for large values of $\tb$.
At lowest order the relation reads $\mb =h_b v_1$. Beyond the tree level
large radiative corrections proportional to $h_b v_2$ are induced,
giving rise to $\tb$-enhanced 
contributions~\cite{deltamb1,deltamb2,deltamb2b,deltamb3}.
At the one-loop level the leading
terms proportional to $v_2$ are generated either by
gluino--sbottom \onel\ diagrams of \order{\als}
or by chargino--stop loops of \order{\alt}. 

The leading one-loop contribution
$\db$ in the limit of $\msusy \gg \mt$ and $\tb \gg 1$ takes the simple
form~\cite{deltamb1}
\BE
\db = \frac{2\als}{3\,\pi} \, \mgl \, \mu \, \tb \,
                    \times \, I(\msbe, \msbz, \mgl) +
      \frac{\alt}{4\,\pi} \, \At \, \mu \, \tb \,
                    \times \, I(\mste, \mstz, \mu) ~,
\label{def:dmb}
\end{equation}
where the function $I$ is given by
\BEA
I(a, b, c) &=& \ed{(a^2 - b^2)(b^2 - c^2)(a^2 - c^2)} \,
               \KL a^2 b^2 \log\frac{a^2}{b^2} +
                   b^2 c^2 \log\frac{b^2}{c^2} +
                   c^2 a^2 \log\frac{c^2}{a^2} \KR \\
 &\sim& \ed{\mbox{max}(a^2, b^2, c^2)} ~. \non
\EEA
The leading contribution can be resummed to all orders in the
perturbative expansion~\cite{deltamb1,deltamb2,deltamb2b}. This leads 
in particular to the replacement 
\BE
\mbms \to \frac{\mbms}{1 + \db} ,
\label{eq:dmbresum}
\end{equation}
where $\mbms$ denotes the running bottom quark mass including SM QCD
corrections. For the numerical evaluations in this paper we choose
$\mbms = \mbms(\mt) \approx 2.97 \gev$. 

The $\db$ corrections are numerically sizable for large $\tb$
in combination with large values of the ratios of $\mu \mgl/ \msusy^2$ or  
$\mu \At/ \msusy^2$. Negative values of $\db$ lead to an enhancement of
the bottom Yukawa coupling as a consequence of \refeq{eq:dmbresum}
(for extreme values of $\mu$ and $\tb$ the bottom Yukawa coupling
can even acquire
non-perturbative values when $\db \to -1$), 
while positive values of $\db$ give rise to a suppression of the 
Yukawa coupling.
Since a change in the sign of $\mu$ reverses the sign of $\db$, the
bottom Yukawa coupling can exhibit a very pronounced dependence on the
parameter $\mu$.

For large values of $\tb$ the correction to the production cross
sections of the Higgs bosons $H$ and $A$ induced by $\db$ enters
approximately like $\TQb/(1 + \db)^2$, giving rise to potentially large
numerical effects. In the case of the subsequent Higgs-boson decay 
$\phi \to \tau^+\tau^-$, however, the $\db$ corrections in the
production and the decay process cancel each other to a large extent. 
The residual $\db$ dependence of 
$\si(b \bar b \phi) \times \br(\phi \to \tau^+\tau^-)$ is 
approximately given by $\TQb/((1 + \db)^2 + 9)$, 
which has a much weaker $\db$ dependence (see \citere{benchmark3} for a
more detailed discussion).

In the numerical analysis below the $\db$ corrections, which have been
discussed in this section in terms of simple approximation formulae,
will be supplemented by other higher-order corrections as implemented in
the program {\tt FeynHiggs} (and possible decay modes into
supersymmetric
particles are taken into account). Higher-order corrections to 
Higgs decays into $\tau^+\tau^-$ within the SM and MSSM have been
evaluated in \citeres{hff,mhiggsf1lC}.


\subsubsection{Corrections to the Higgs production cross sections}
\label{sec:HiggsXS}

For the prediction of
Higgs-boson production processes at hadron colliders 
SM-type QCD corrections in general play an important role. 
The SM predictions for the process
$b\bar b \to \phi + X$ at the LHC are far advanced.
In the five-flavor scheme the SM cross section
is known at NNLO in QCD~\cite{Harlander:2003ai}.  The cross section in the
four-flavor scheme is known at
NLO~\cite{Dittmaier:2003ej,Dawson:2003kb}.  Results obtained in the
two schemes have been shown to be
consistent~\cite{Assamagan:2004mu,Dawson:2004sh,Dawson:2005vi}
(see also \citeres{Campbell:2002zm,Dawson:2004sh} and
\citeres{Dittmaier:2003ej,Dawson:2003kb} for results with one and two
final-state $b$-quarks at high-$p_T$, respectively). 

The predictions for the $b\bar b \to \phi + X$ cross sections in the
MSSM have been obtained with 
{\tt FeynHiggs}~\cite{feynhiggs,mhiggslong,mhiggsAEC,feynhiggs2.5}. 
The {\tt FeynHiggs} implementation%
\footnote{The inclusion of the charged Higgs production cross sections is
planned for the near future.}
is based on the state-of-the-art SM prediction, 
namely the NNLO result in the five-flavor scheme~\cite{Harlander:2003ai} 
using MRST2002 parton distributions at NNLO~\cite{Martin:2002aw},
with the renormalization scale set 
equal to $\MHSM$ and the factorization scale set equal to $\MHSM/4$.
In order to obtain the MSSM prediction the SM cross section
is rescaled with the ratio of the partial widths in the MSSM and the SM,
\BE
\frac{\Ga(\phi \to b \bar b)_{\MSSM}}{\Ga(\phi \to b \bar b)_{\SM}}~.
\EE
The evaluation of the partial widths incorporates
one-loop SM QCD and SUSY QCD corrections, as well as (in the SUSY case)
the resummation of all terms of 
\order{(\al_s \tb)^n}~\cite{hff,mhiggsf1lC,deltamb2} and the proper
normalization of the external Higgs bosons as discussed in
\citeres{Hahn:2002gm,feynhiggs2.5}.
Since the approximation of rescaling the SM cross
section with the ratio of partial widths does not take into account the
MSSM-specific dynamics of the production processes, the theoretical
uncertainty in the predictions for the cross sections will in
general be somewhat larger than for the decay widths.
It should be noted that in comparison with other approaches for treating
the SM and SUSY contributions, for instance the program 
{\tt HQQ}~\cite{hqq}, sizable deviations can occur as a consequence of
differences in the scale choices and the inclusion of higher-order
corrections.


\subsection{The $\mhmax$ and no-mixing benchmark scenarios}
\label{sec:benchmarks}

While the phenomenology of the production and decay processes of the
heavy neutral MSSM Higgs bosons at the LHC is mainly characterised by
the parameters $\MA$ and $\tb$ that govern the Higgs sector at lowest
order, other MSSM parameters enter via higher-order contributions, as 
discussed above, and via the kinematics of Higgs-boson decays into
supersymmetric particles. The other MSSM parameters are usually fixed
in terms of benchmark scenarios. The most commonly used scenarios are
the ``$\mhmax$'' and ``no-mixing'' benchmark 
scenarios~\cite{benchmark,benchmark2,benchmark3}. According to the
definition of \citere{benchmark2} the $\mhmax$ scenario is given by
\BEA
\mbox{\underline{$\mhmax:$}} &&
\msusy = 1000 \gev, \quad \Xt = 2\, \msusy, \quad \Ab = \At, \non \\
&& \mu = 200 \gev, \quad M_2 = 200 \gev, \quad \mgl = 0.8\,\msusy~.
\label{mhmax}
\EEA
The no-mixing scenario differs from the $\mhmax$ scenario only in that it has 
vanishing mixing in the stop sector and a larger value of $\msusy$
\BEA
\mbox{\underline{no-mixing:}} &&
\msusy = 2000 \gev, \quad \Xt = 0, \quad \Ab = \At, \non \\
&& \mu = 200 \gev, \quad M_2 = 200 \gev, \quad \mgl = 0.8\,\msusy~.
\label{nomix}
\EEA
The value of the top-quark mass in \citere{benchmark2} was chosen
according to the experimental central value at that time. For our
numerical analysis below, we use 
the value, $\mt = 171.4 \gev$~\cite{mt1714}%
\footnote{
Most recently the central experimental value has shifted to
$\mt = 170.9 \pm 1.8 \gev$~\cite{mt1709}. This shift has a negligible impact
on our analysis.
}%
.

In \citere{benchmark3} it was suggested that in the search for heavy
MSSM Higgs bosons the $\mhmax$ and no-mixing scenarios, which originally
were mainly designed for the search for the light $\cp$-even Higgs boson
$h$, should be extended by several discrete values of $\mu$,
\BE
\mu = \pm 200, \pm 500, \pm 1000 \gev ~.
\label{eq:variationmu}
\EE
As discussed above, the variation of $\mu$ in particular has an impact
on the correction $\db$, modifying in this way the bottom Yukawa
coupling. For very large values of $\tb$ and
large negative values of $\mu$ the bottom Yukawa coupling can be so much 
enhanced that a perturbative treatment is no longer possible. We
have checked that in our analysis of the LHC discovery contours 
the bottom Yukawa coupling stays in the perturbative regime, so that
all values of $\mu$ down to $\mu = -1000 \gev$ can safely be inserted.

The variation of the parameter $\mu$ also modifies the mass spectrum and
the couplings in the chargino and neutralino sector of the MSSM. Besides
the small higher-order corrections induced by loop diagrams involving
charginos and neutralinos, a change in the mass spectrum of the chargino
and neutralino sector can have an important effect on Higgs phenomenology
because decay modes of the heavy neutral MSSM Higgs bosons into
charginos and neutralinos open up if the supersymmetric particles
are sufficiently light
(the mass spectrum in the $\mhmax$ and no-mixing scenarios respects the 
limits from direct searches for charginos at LEP~\cite{pdg} for all
values of $\mu$ specified in \refeq{eq:variationmu}).

Differences between the $\mhmax$ and no-mixing scenarios in the searches
for heavy neutral MSSM Higgs bosons are induced in particular by a
difference in the $\db$ correction. While in the $\mhmax$ scenario
both the \order{\als} and \order{\alt} contributions to $\db$
can be sizable, see \refeq{def:dmb}, in the no-mixing scenario the
\order{\alt} contribution is very small because $\At$ is close 
to zero in this case. The larger value of $\msusy$ in the no-mixing
scenario gives rise to an additional suppression of $|\db|$ compared to
the $\mhmax$ scenario.


\section{Experimental analysis}
\label{sec:expanal}

In this section we briefly review the recent CMS analysis of the
$\phi \to \tau^+\tau^-$ channel, see \citere{lhctdrsS}, 
yielding the number of events
needed for a 5$\,\si$ discovery (depending on the mass of the Higgs
boson). The analysis was performed with full CMS detector 
simulation and reconstruction for the following
four final states of di-$\tau$-lepton decays:  
$\tau^+\tau^- \to \,\mbox{jets}$~\cite{CMSPTDRjj},  
$\tau^+\tau^- \to e + \,\mbox{jet}$~\cite{CMSPTDRej}, 
$\tau^+\tau^- \to \mu + \mbox{jet}$~\cite{CMSPTDRmj}
and $\tau^+\tau^- \to e + \mu$~\cite{CMSPTDRem}. 

The Higgs-boson production in 
association with $b$ quarks, $pp \to b\bar b  \phi$, has been selected using
single $b$-jet tagging in the experimental analysis. The kinematics of the 
$gg \to b\bar b \phi$ production process (2 $\to$ 3) was generated with 
PYTHIA~\cite{PYTHIA}. It
has been shown that in this way the NLO kinematics is better reproduced than
using the PYTHIA $gb \to b\phi$ process (2 $\to$ 2)~\cite{bbh}. 
The backgrounds considered in the analysis were QCD muli-jet events 
(for the $\tau \tau \to \,\mbox{jets}$ mode), 
$t \bar t, b \bar b$,  Drell-Yan production of $Z, \ga^{\ast}$, $W$+jet, $Wt$
and $\tau \tau b \bar b$. All background processes  
were generated using PYTHIA, except for $\tau^{+} \tau^{-} b \bar b$,
which was generated using CompHEP \cite{Boos:2004kh}.

\begin{table}[htb!]
\renewcommand{\arraystretch}{1.5}
\BC
\begin{tabular}{|c||c|c|c|} \hline
\multicolumn{4}{|c|}
  {$\phi \to \tau^+\tau^- \to \,\mbox{jets}$, 60~\ifb} \\ \hline\hline
$\MA$ [GeV]     & 200   & 500   & 800 \\ \hline
$N_S$           &  63   &  35   &  17 \\ \hline
$\eps_{\rm exp}$ & $2.5 \times 10^{-4}$ & $2.4 \times 10^{-3}$ & 
                                        $3.6 \times 10^{-3}$ \\ \hline
$R_{M_\phi}$     & 0.176 & 0.171 & 0.187 \\ \hline
$\De M_{\phi} / M_{\phi}$ [\%] & 2.2 & 2.8 & 4.5 \\
\hline
\end{tabular}
\EC
\vspace{-1em}
\caption{
Required number of signal events, $N_S$, with $\cL = 60$~\ifb\ for a 5$\,\si$
discovery in the channel $\phi \to \tau^+\tau^- \to \,\mbox{jets}$. 
Furthermore given are the
total experimental selection efficiency, $\eps_{\rm exp}$, the ratio of
the di-$\tau$ mass resolution to the Higgs-boson mass, $R_{M_\phi}$, 
and the expected precision of the Higgs-boson mass measurement,
$\De M_{\phi} / M_{\phi}$, obtainable from $N_S$ signal events.
}
\label{tab:jj}
\renewcommand{\arraystretch}{1.0}
\end{table}

\begin{table}[htb!]
\vspace{1em}
\renewcommand{\arraystretch}{1.5}
\BC
\begin{tabular}{|c||c|c|c|} \hline
\multicolumn{4}{|c|}
  {$\phi \to \tau^+\tau^- \to e + \,\mbox{jet}$, 30~\ifb} \\ \hline\hline
$\MA$ [GeV]      & 200   & 300   & 500   \\ \hline
$N_S$            &  72.9 &  45.5 &  32.8 \\ \hline
$\eps_{\rm exp}$  & $3.0 \times 10^{-3}$ & $6.4 \times 10^{-3}$ & 
                                           $1.0 \times 10^{-2}$ \\ \hline
$R_{M_\phi}$      & 0.216 & 0.214 & 0.230 \\ \hline
$\De M_{\phi} / M_{\phi}$ [\%] & 2.5 & 3.2 & 4.0 \\ \hline
\end{tabular}
\EC
\vspace{-1em}
\caption{
Required number of signal events, $N_S$, with $\cL = 30$~\ifb\ for 
a 5$\,\si$ discovery in the channel
$\phi \to \tau^+\tau^- \to e + \,\mbox{jet}$. 
The other quantities are defined as in \refta{tab:jj}.
}
\label{tab:ej}
\renewcommand{\arraystretch}{1.0}
\end{table}

\begin{table}[htb!]
\vspace{1em}
\renewcommand{\arraystretch}{1.5}
\BC
\begin{tabular}{|c||c|c|} \hline
\multicolumn{3}{|c|}
  {$\phi \to \tau^+\tau^- \to \mu + \,\mbox{jet}$, 30~\ifb} \\ \hline\hline
$\MA$ [GeV]     & 200  & 500  \\ \hline
$N_S$           &  79  &  57  \\ \hline
$\eps_{\rm exp}$ & $7.0 \times 10^{-3}$ & $2.0 \times 10^{-2}$ \\ \hline
$R_{M_\phi}$     & 0.210 & 0.200 \\ \hline
$\De M_{\phi} / M_{\phi}$ [\%] & 2.4 & 2.6 \\ \hline
\end{tabular}
\EC
\vspace{-1em}
\caption{
Required number of signal events, $N_S$, with $\cL = 30$~\ifb\ for a 5$\,\si$
discovery in the channel $\phi \to \tau^+\tau^- \to \mu + \,\mbox{jet}$.
The other quantities are defined as in \refta{tab:jj}.
}
\label{tab:mj}
\renewcommand{\arraystretch}{1.0}
\end{table}

\begin{table}[htb!]
\renewcommand{\arraystretch}{1.5}
\BC
\begin{tabular}{|c||c|c|} \hline
\multicolumn{3}{|c|}
  {$\phi \to \tau^+\tau^- \to e + \mu$, 30~\ifb} \\ \hline\hline
$\MA$ [GeV]     & 200    & 250 \\ \hline
$N_S$           & 87.8  & 136.7 \\ \hline
$\eps_{\rm exp}$ & $6.4 \times 10^{-3}$ & $1.1 \times 10^{-2}$ \\ \hline
$R_{M_\phi}$     & 0.262 & 0.412 \\ \hline
$\De M_{\phi} / M_{\phi}$ [\%] & 2.8 & 3.5 \\ \hline
\end{tabular}
\EC
\caption{
Required number of signal events, $N_S$, with $\cL = 30$~\ifb\ for a 5$\,\si$
discovery in the channel $\phi \to \tau^+\tau^- \to e + \mu$. 
The other quantities are defined as in \refta{tab:jj}.
}
\label{tab:em}
\renewcommand{\arraystretch}{1.0}
\end{table}

The results for the various channels, \refeqs{jj} -- (\ref{em}), are given in
\reftas{tab:jj} -- \ref{tab:em}. For every Higgs-boson mass point studied
we show the number of signal events needed for 5$\,\si$ discovery, $N_S$, the
total experimental selection efficiency, $\eps_{\rm exp}$, and the ratio of
the di-$\tau$ mass resolution to the Higgs-boson mass, $R_{M_\phi}$. The
last row in \reftas{tab:jj} -- \ref{tab:em} shows the expected 
precision of the Higgs-boson mass measurement, evaluated as explained
below, for parameter points on the 5$\,\si$ discovery contour.
Detector effects, experimental systematics and uncertainties
of the background determination were taken into account in the evaluation of
the $N_S$. These effects reduce
the discovery region in the $\MA$--$\tb$ plane as shown
in previous analyses~\cite{lhctdrsS} (see in particular
Fig.~5.6 of \citere{lhctdrsS} for the 
$\tau^+\tau^- \to \mu + \,\mbox{jet}$ mode).


Now we turn to the evaluation of the expected precision of the 
Higgs-boson mass measurement. 
In spite of the escaping neutrinos, the Higgs-boson mass can be 
reconstructed in the $H,A \to \tau \tau$ channel from the visible $\tau$ 
momenta ($\tau$ jets) and the missing transverse energy, 
$E_{\rm T}^{\rm miss}$, using the collinearity approximation for neutrinos from
highly boosted $\tau$'s. 
In the investigated region of $\MA$ and $\tb$ the two states $A$ and $H$ are  
nearly mass-degenerate.
For most values of the other MSSM parameters the mass difference of $A$
and $H$ is much smaller than the achievable mass resolution. In this
case the difference in reconstructing the $A$ or the $H$ will have no
relevant effect on the achievable accuracy in the mass determination.
In some regions of the MSSM parameter space, however, a sizable
splitting between $\MA$ and $\MH$ can occur even for $\MA \gg \MZ$.
We will discuss below the prospects in scenarios where the 
splitting between $\MA$ and $\MH$ is relatively large.
The precision $\De M_\phi/M_\phi$ shown in \reftas{tab:jj} -- \ref{tab:em}
is derived for the border of the parameter space in
which a 5$\,\si$ discovery can be claimed, i.e.\ with $N_S$ observed Higgs
events. The statistical accuracy of the mass measurement has been evaluated
via 
\BE
\label{eq:precM}
\frac{\De M_\phi}{M_\phi} = \frac{R_{M_\phi}}{\sqrt{N_S}}~.
\EE
A higher precision can be achieved if more than $N_S$ events are observed. The
corresponding estimate for the precision is obtained 
by replacing $N_S$ in \refeq{eq:precM} by the number of observed
signal events, $N_{\rm ev}$. It should be noted that the prospective 
accuracy obtained from \refeq{eq:precM} does not take into account 
the uncertainties of the jet and missing $E_{\rm T}$ energy scales. In 
the $\tau^+\tau^- \to \,\mbox{jets}$ mode these effects
can lead to an additional 3\% uncertainty in the mass 
measurement~\cite{CMSPTDRjj}.
A more dedicated 
procedure of the mass measurement from the signal plus background data 
still has to be developed in the experimental analysis. However, we do not
expect that the additional uncertainties will considerably degrade the
accuracy of the Higgs boson mass measurement as calculated with
\refeq{eq:precM}.


\section{Results}
\label{sec:results}

The results 
quoted in \refse{sec:expanal} for the required number of signal events
depend only on the Higgs-boson
mass, i.e.\ the event kinematics,
but are independent of any specific MSSM scenario. 
In order to determine the 5$\,\si$ discovery contours in the
$\MA$--$\tb$~plane these results have to be confronted with the MSSM
predictions.
The number of signal events, $N_{\rm ev}$, for a given parameter
point is evaluated via
\BE
N_{\rm ev} = \cL \times \si_{b\bar b\phi} \times 
             \br(\phi \to \tau^+\tau^-) \times \br_{\tau\tau} \times
             \eps_{\rm exp}~.
\EE
Here $\cL$ denotes the luminosity collected with the CMS detector, 
$\si_{b\bar b\phi}$ is the Higgs-boson production cross section, 
$\br(\phi \to \tau^+\tau^-)$ is the branching ratio of
the Higgs boson to $\tau$~leptons, 
$\br_{\tau\tau}$ is the product of the branching ratios of the two
$\tau$~leptons into their respective final state, 
\BEA
\br(\tau \to \,\mbox{jet} + X) &\approx& 0.65~, \\
\br(\tau \to \mu + X) \approx \br(\tau \to e + X) &\approx& 0.175~,
\EEA
and $\eps_{\rm exp}$ denotes the total experimental selection efficiency
for the respective
process (as given in \reftas{tab:jj} -- \ref{tab:em}). The Higgs-boson
production cross sections and decay branching ratios have been evaluated
with {\tt FeynHiggs} as described in \refse{sec:higherorder}.


\subsection{Discovery reach for heavy neutral MSSM Higgs bosons}

The number of signal events, $N_{\rm ev}$, in the MSSM depends besides
the parameters $\MA$ and $\tb$, which govern the MSSM Higgs sector at
lowest order, in principle also on all other MSSM parameters. In the
following we analyze how stable the results for the $5 \si$ discovery
contours in the $\MA$--$\tb$ plane are with respect to variations of
the other MSSM parameters. We take into account both effects from
higher-order corrections, as discussed in \refse{sec:higherorder}, and 
from decays of the heavy Higgs bosons into supersymmetric particles.
As starting point of our analysis we use the $\mhmax$ and no-mixing
benchmark scenarios, where we investigate in detail the sensitivity of the 
discovery contours with respect to variations of the parameter $\mu$.
We then discuss the possible impact of varying other MSSM parameters.

We have evaluated $N_{\rm ev}$ in the two benchmark scenarios as a
function of $\MA$ and $\tb$.
For fixed $\MA$ we have varied $\tb$ such that $N_{\rm ev} = N_S$ (as given in
\reftas{tab:jj} -- \ref{tab:em}). This $\tb$ value is then identified as 
the point on the 5$\,\si$ discovery contour corresponding to the chosen
value of $\MA$.
In this way we have determined the 5$\,\si$ discovery contours
for the $\mhmax$ and the no-mixing scenarios for 
$\mu = \pm 200, \pm 1000 \gev$. 

\begin{figure}[htb!]
\vspace{1em}
\BC
\includegraphics[width=.49\textwidth]{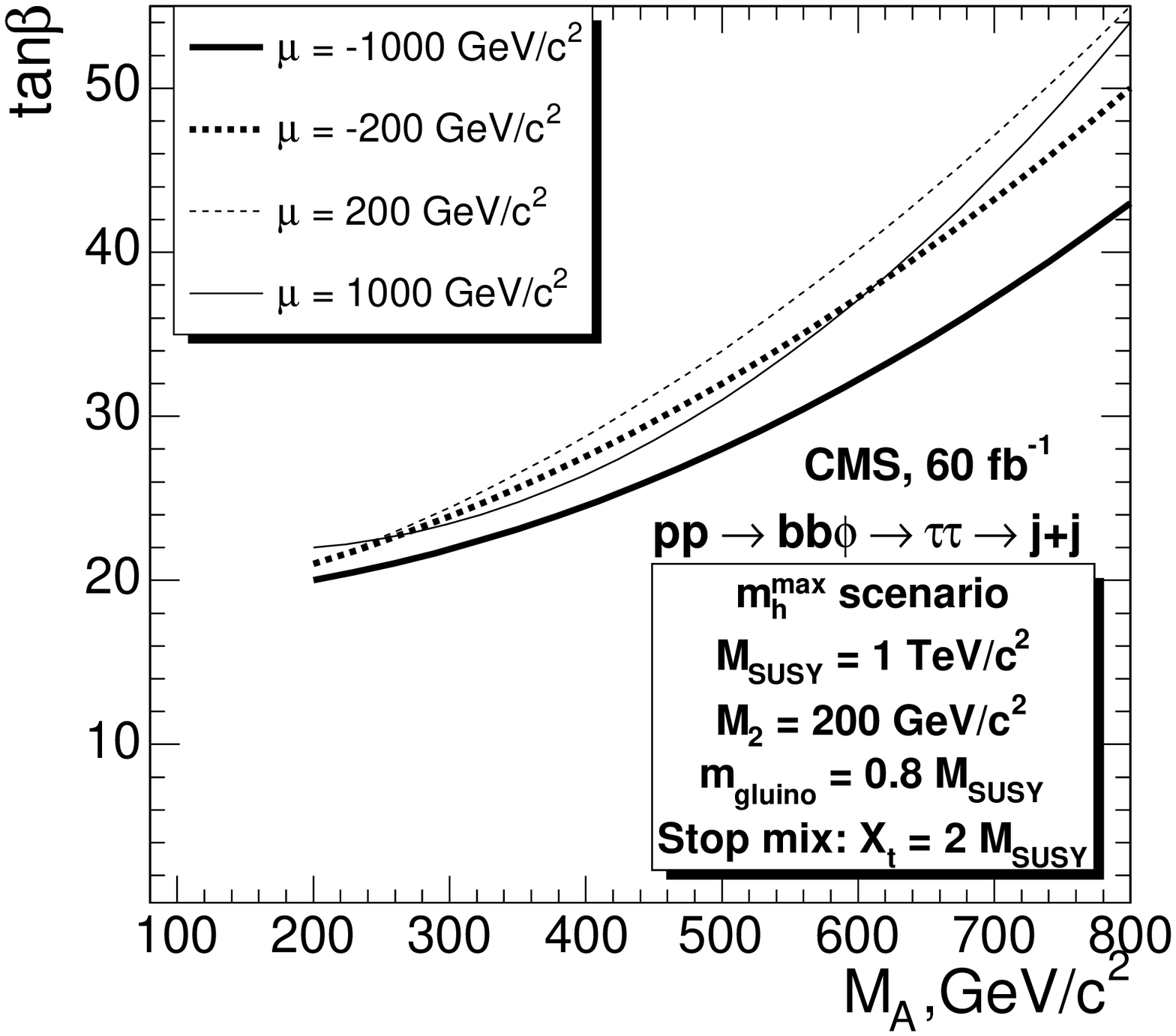}
\includegraphics[width=.49\textwidth]{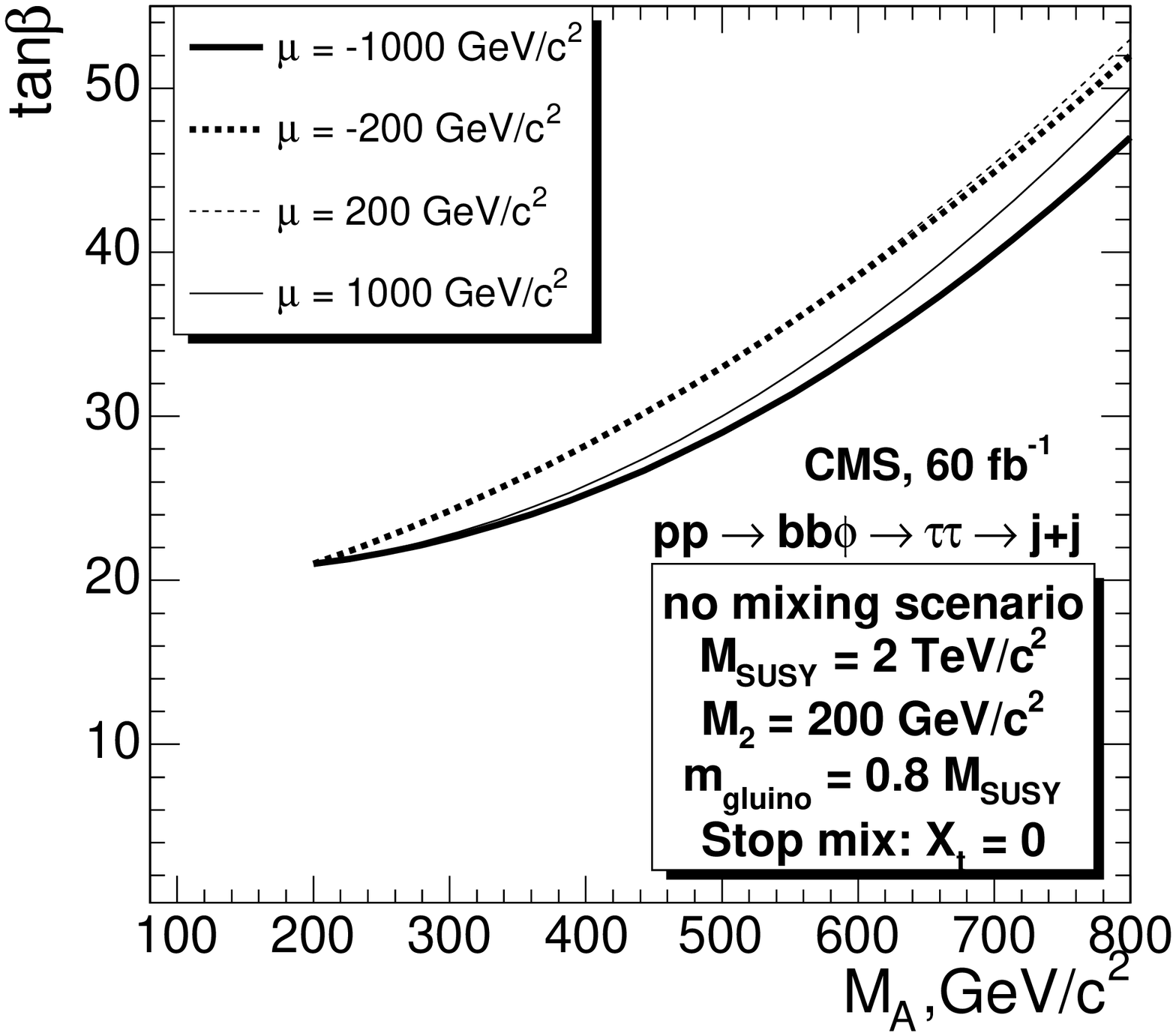}
\EC
\caption{Variation of the $5 \si$ discovery contours obtained 
from the channel $b\bar b \phi, \phi \to \tau^+\tau^- \to \,\mbox{jets}$
in the $\mhmax$ (left) and no-mixing (right) benchmark scenarios 
for different values of~$\mu$.}
\label{fig:jj}
\vspace{2em}
\end{figure}

\begin{figure}[htb!]
\vspace{1em}
\BC
\includegraphics[width=.49\textwidth]{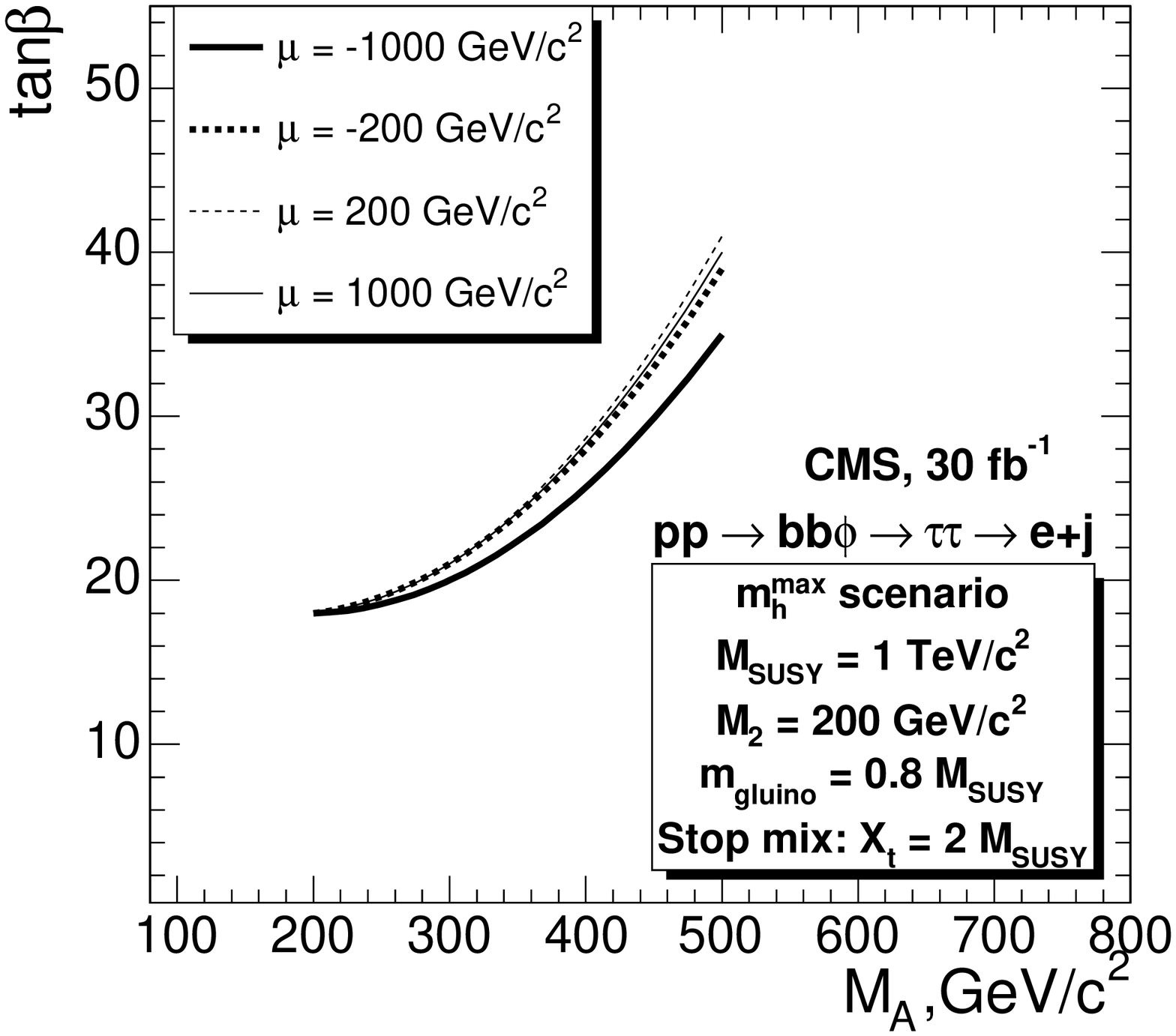}
\includegraphics[width=.49\textwidth]{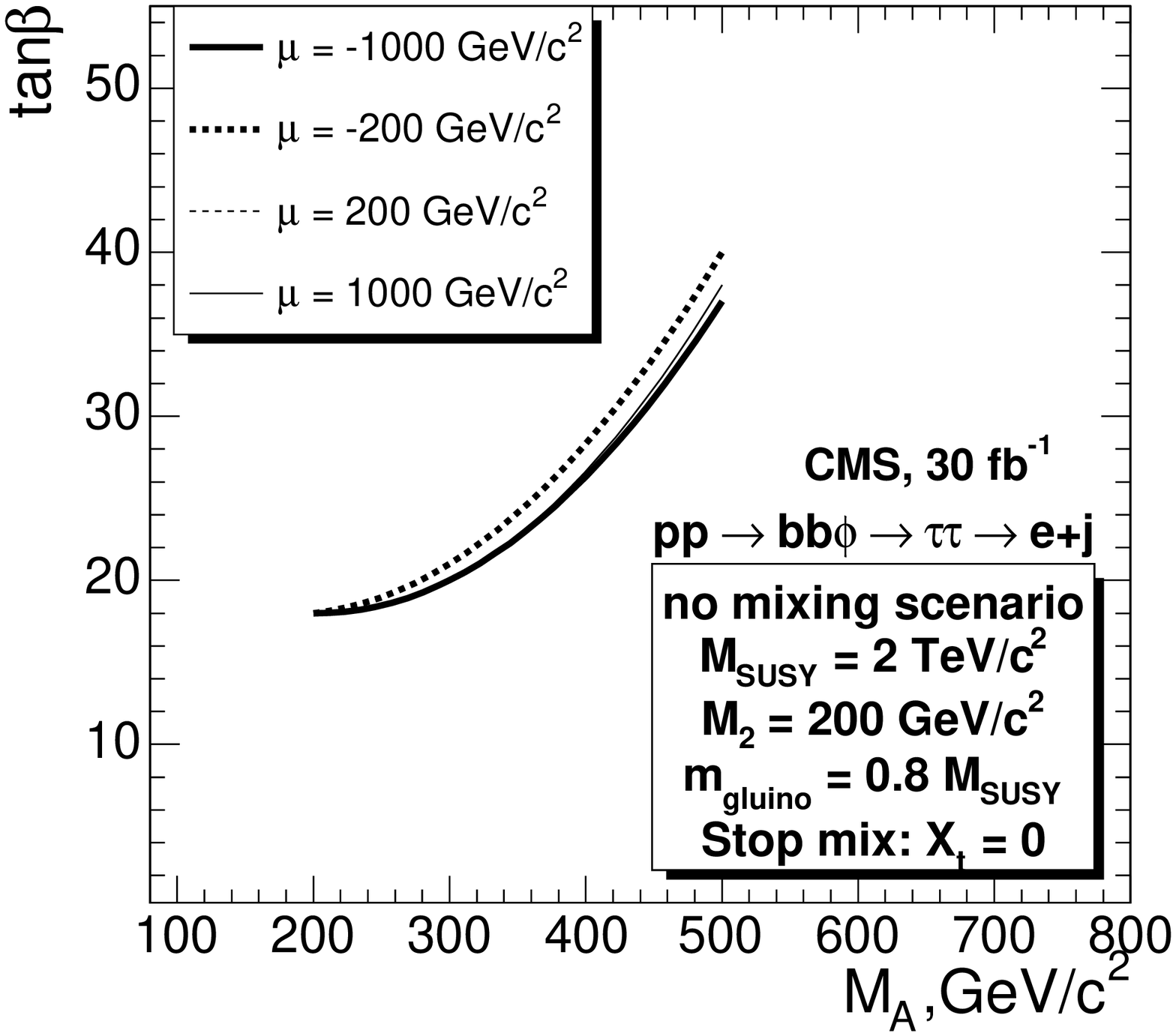}
\EC
\caption{Variation of the $5 \si$ discovery contours obtained 
from the channel $b\bar b \phi, \phi \to \tau^+\tau^- \to e + \,\mbox{jet}$
in the $\mhmax$ (left) and no-mixing (right) benchmark scenarios 
for different values of~$\mu$.}
\label{fig:ej}
\vspace{2em}
\end{figure}

\begin{figure}[htb!]
\vspace{1em}
\BC
\includegraphics[width=.49\textwidth]{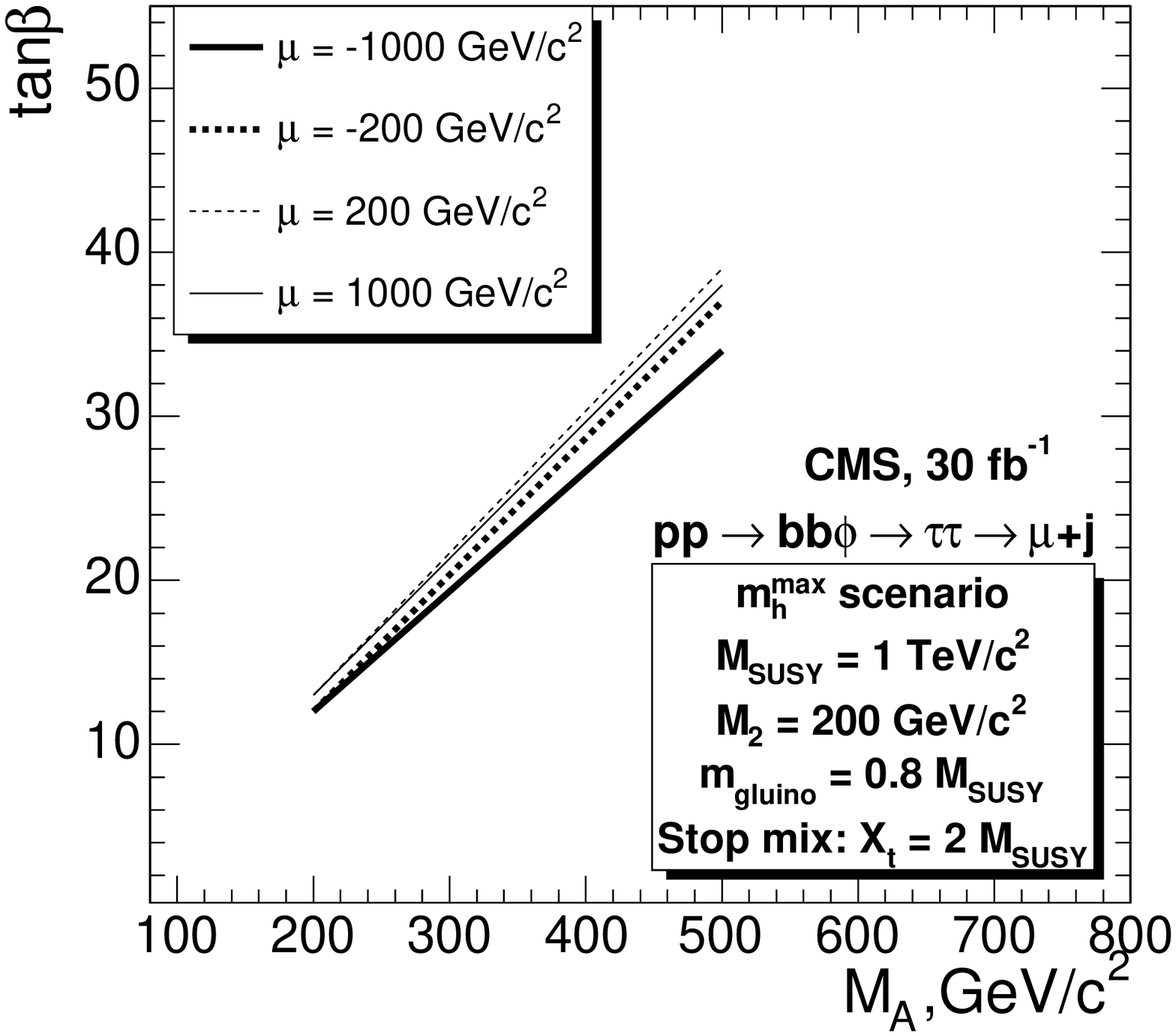}
\includegraphics[width=.49\textwidth]{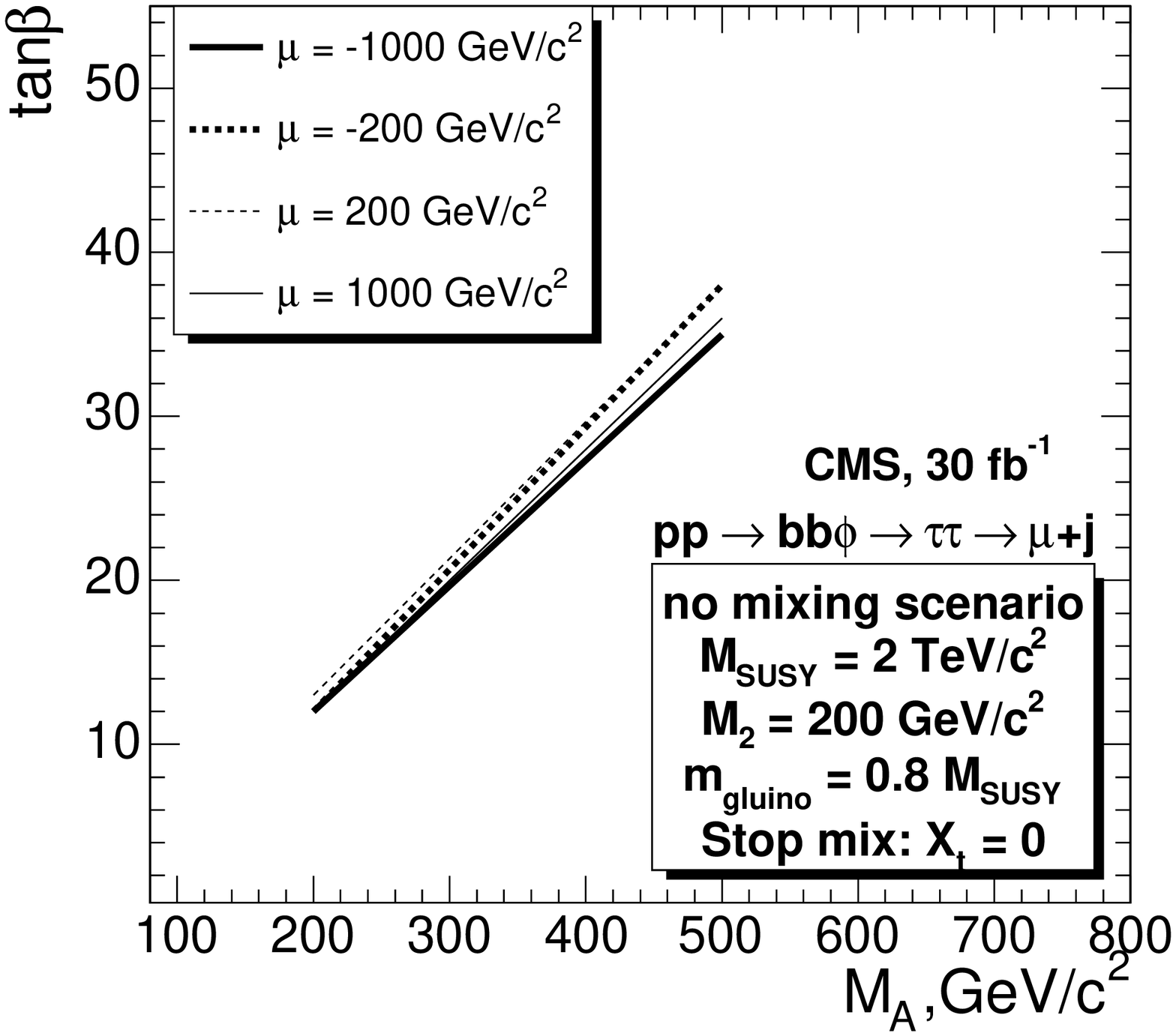}
\EC
\caption{Variation of the $5 \si$ discovery contours obtained 
from the channel $b\bar b \phi, \phi \to \tau^+\tau^- \to \mu + \,\mbox{jet}$
in the $\mhmax$ (left) and no-mixing (right) benchmark scenarios 
for different values of~$\mu$.}
\label{fig:mj}
\vspace{2em}
\end{figure}

In \reffis{fig:jj} -- \ref{fig:mj} we show 
the $5 \si$ discovery contours obtained from the process 
$b\bar b \phi, \phi \to \tau^+\tau^-$ for the final states
$\tau^+\tau^- \to \,\mbox{jets}$,  
$\tau^+\tau^- \to e + \,\mbox{jet}$ and 
$\tau^+\tau^- \to \mu + \,\mbox{jet}$.
As can be seen from \refta{tab:em}, the fourth channel discussed above,
$\tau^+\tau^- \to e + \mu$, contributes for 30~fb$^{-1}$ only in the
region of relatively small $\MA$ values 
and has a lower sensitivity than the other three channels. We
therefore omit this channel in the following discussion.
The discovery contours in \reffis{fig:jj} -- \ref{fig:mj} are given 
for the $\mhmax$ and no-mixing benchmark scenarios with 
$\mu = \pm 200, \pm 1000 \gev$. 
As explained above, the 5$\,\si$ discovery contours are
affected by a change in $\mu$ in two ways. Higher-order contributions,
in particular the ones associated with $\db$, modify the Higgs-boson
production cross sections and decay branching ratios. 
Furthermore the mass eigenvalues of the charginos and neutralinos vary
with $\mu$, possibly opening up the decay channels of the Higgs bosons
to supersymmetric
particles, which reduces the branching ratio to $\tau$~leptons. 

The results for the 5$\,\si$ discovery contours for the final state 
$\tau^+\tau^- \to \,\mbox{jets}$ are shown in \reffi{fig:jj} for 
the $\mhmax$ (left) and the no-mixing (right) scenario. 
As expected from the discussion of the $\db$ corrections in
\refse{sec:higherorder},
the variation of the 5$\,\si$ discovery contours with $\mu$
is more pronounced in the $\mhmax$ scenario, where a shift up to 
$\De\tb = 12$ can be observed for 
$\MA = 800 \gev$. For low $\MA$ values (corresponding also to lower
$\tb$ values on the discovery contours) the variation stays below
$\De\tb = 3$. In the no-mixing scenario the variation does not exceed 
$\De\tb = 5$. 
The $\tau^+\tau^- \to \,\mbox{jets}$ channel has also been discussed in
\citere{benchmark3}. Our results, which are based on the latest CMS
studies using full simulation~\cite{CMSPTDRjj}, are qualitatively in good
agreement with \citere{benchmark3}, in which the earlier CMS studies of 
\citeres{cmshiggs,KinNik} had beed used.
The 5$\,\si$ discovery regions are
largest for $\mu = -1000 \gev$ and pushed to highest $\tb$ values for
$\mu = +200 \gev$. In the low $\MA$ region our discovery contours are
very similar to those obtained in \citere{benchmark3}. In the high $\MA$
region, $\MA \sim 800 \gev$, corresponding to larger values of
$\tb$ on the discovery contours, our improved evaluation of the 5$\,\si$ 
discovery contours gives rise to a shift towards higher $\tb$ values 
compared to \citere{benchmark3} of about $\De\tb = 8$ (mostly due to
the up-to-date experimental input).
Accordingly, we find a smaller discovery region compared to
\citere{benchmark3} and therefore an enlarged
``LHC wedge'' region where only the light $\cp$-even  MSSM Higgs boson 
can be detected at the 5$\,\si$ level.

The results for the channel $\tau^+\tau^- \to e + \,\mbox{jet}$ are shown in
\reffi{fig:ej}. Again the $\mhmax$ scenario shows a stronger variation than
the no-mixing scenario. The resulting 
shift in $\tb$ reaches up to $\De\tb = 8$ for $\MA = 500 \gev$ in the
$\mhmax$ scenario, but stays below $\De\tb = 4$ for the no-mixing
scenario. 
Finally in \reffi{fig:mj} the results for the channel 
$\tau^+\tau^- \to \mu + \,\mbox{jet}$ are depicted. The level of variation of
the 5$\,\si$ discovery contours is the same as for the $e + \,\mbox{jet}$
final state.%
\footnote{Since the results of the experimental simulation for this
channel are available only for two $\MA$ values,
the interpolation is a straight line. This may result in a slightly larger
uncertainty of the results shown in \reffi{fig:mj} compared to the other
figures.} 
%

\begin{figure}[htb!]
\vspace{1em}
\BC
\includegraphics[width=.49\textwidth]{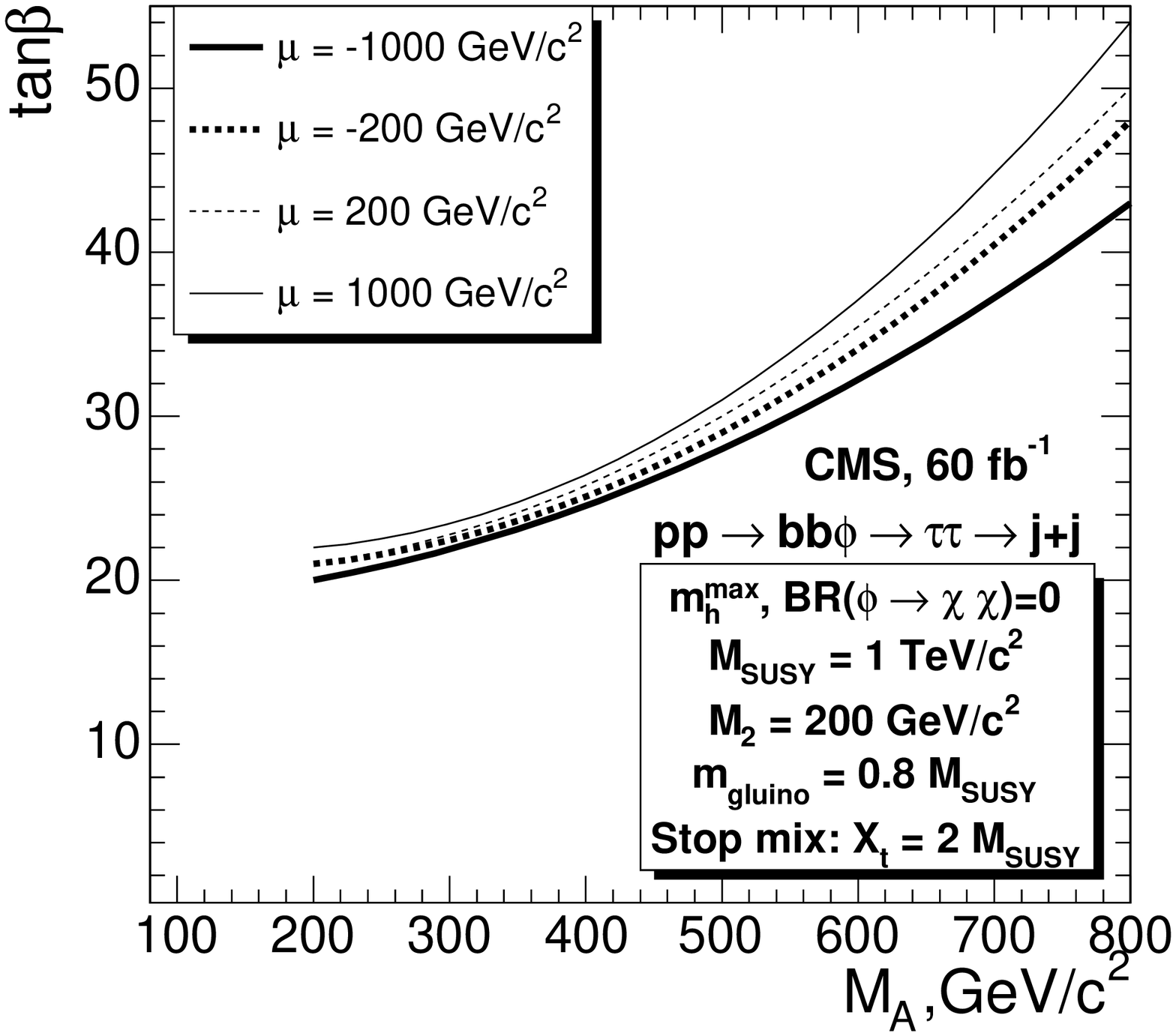}
\includegraphics[width=.49\textwidth]{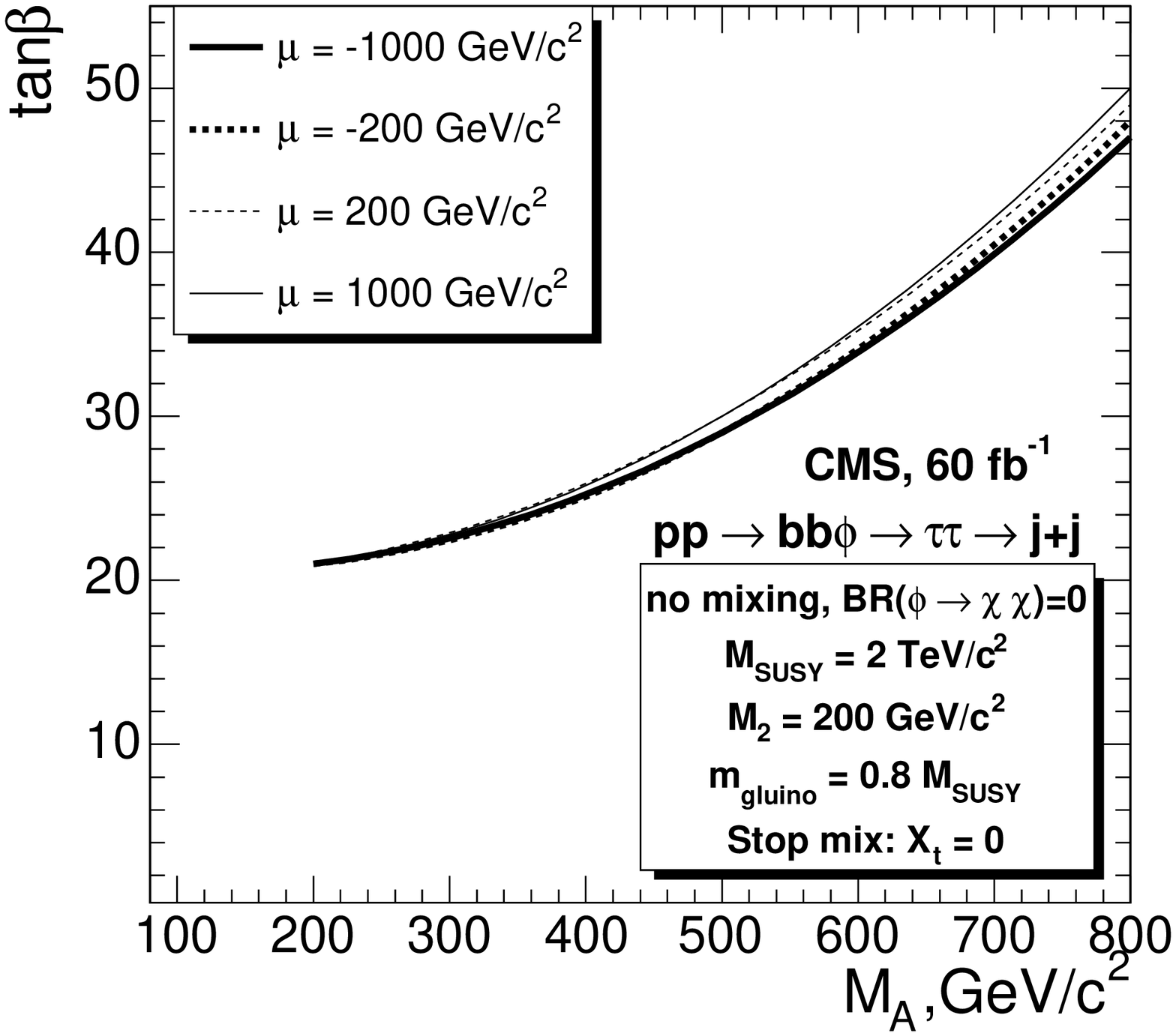}
\EC
\caption{Variation of the $5 \si$ discovery contours obtained 
from the channel $b\bar b \phi, \phi \to \tau^+\tau^- \to \,\mbox{jets}$
in the $\mhmax$ (left) and no-mixing (right) benchmark scenarios 
for different values of~$\mu$ in the case where no decays of the heavy
Higgs bosons into supersymmetric particles are taken into account (see
text).}
\label{fig:jj_brchi0}
\vspace{2em}
\end{figure}

In order to gain a better understanding of how sensitively 
the discovery contours in the $\MA$--$\tb$ plane depend on the chosen
SUSY scenario, it is useful to separately investigate the different effects
caused by varying the parameter $\mu$. For simplicity, we restrict
the following discussion to the 
$b\bar b \phi, \phi \to \tau^+\tau^- \to \,\mbox{jets}$ channel.
In \reffi{fig:jj_brchi0} we show the same results as in \reffi{fig:jj},
but for the case where no decays of the heavy Higgs bosons into
supersymmetric particles are taken into account. As a consequence, the 
variation of the 5$\,\si$ discovery contours with $\mu$ shown in
\reffi{fig:jj_brchi0} is purely an effect of higher-order corrections, 
predominantly those entering via $\db$. The difference between 
\reffi{fig:jj} and \reffi{fig:jj_brchi0}, on the other hand, is purely
an effect of the change in $\br(\phi \to \tau^+\tau^-)$ caused by the
variation of the partial Higgs-boson decay widths into supersymmetric
particles arising from a shift in the masses of the charginos and
neutralinos.

In \reffi{fig:jj_brchi0} the dependence of the 5$\,\si$ discovery contours 
on $\mu$ significantly differs from the case of \reffi{fig:jj}. While
in \reffi{fig:jj}
the inclusion of decays into supersymmetric particles gives rise to the
fact that the smallest discovery region is found for small $\mu$ values,
$\mu = +200 \gev$ (with the exception of the region of very small
$\MA$), in \reffi{fig:jj_brchi0} the 5$\,\si$ discovery contours 
are ordered monotonously in $\mu$: the largest (smallest) 5$\,\si$
discovery regions are obtained for $\mu = -(+)1000 \gev$, i.e.\ for the
largest (smallest) values of the bottom Yukawa coupling. As expected,
the effect of the higher-order corrections is largest in the high
$\tb$-region (corresponding to large values of $\MA$ on the discovery
contours). In this region the variation of $\mu$ shifts the discovery
contours by up to $\De\tb = 11$ for the case of the $\mhmax$ scenario 
(left plot of \reffi{fig:jj_brchi0}), 
i.e.\ the effect is about the same as
for the case where decays into supersymmetric particles are included.
For lower values of $\tb$ (corresponding to smaller values of $\MA$ on 
the discovery contours), on the other hand, the modification of the
Higgs branching ratio as a consequence of decays into supersymmetric
particles yields the dominant effect on the 5$\,\si$ discovery contours.
Accordingly, the observed variation with $\mu$ in this region is
significantly smaller in \reffi{fig:jj_brchi0} as compared to the full
result of \reffi{fig:jj}. The reduced sensitivity of the discovery
contours on $\mu$ can also clearly be seen for the case of the no-mixing
scenario (right plot), where as discussed above the $\db$ correction is
smaller than in the $\mhmax$ scenario.

\begin{figure}[htb!]
\BC
\includegraphics[width=.49\textwidth]{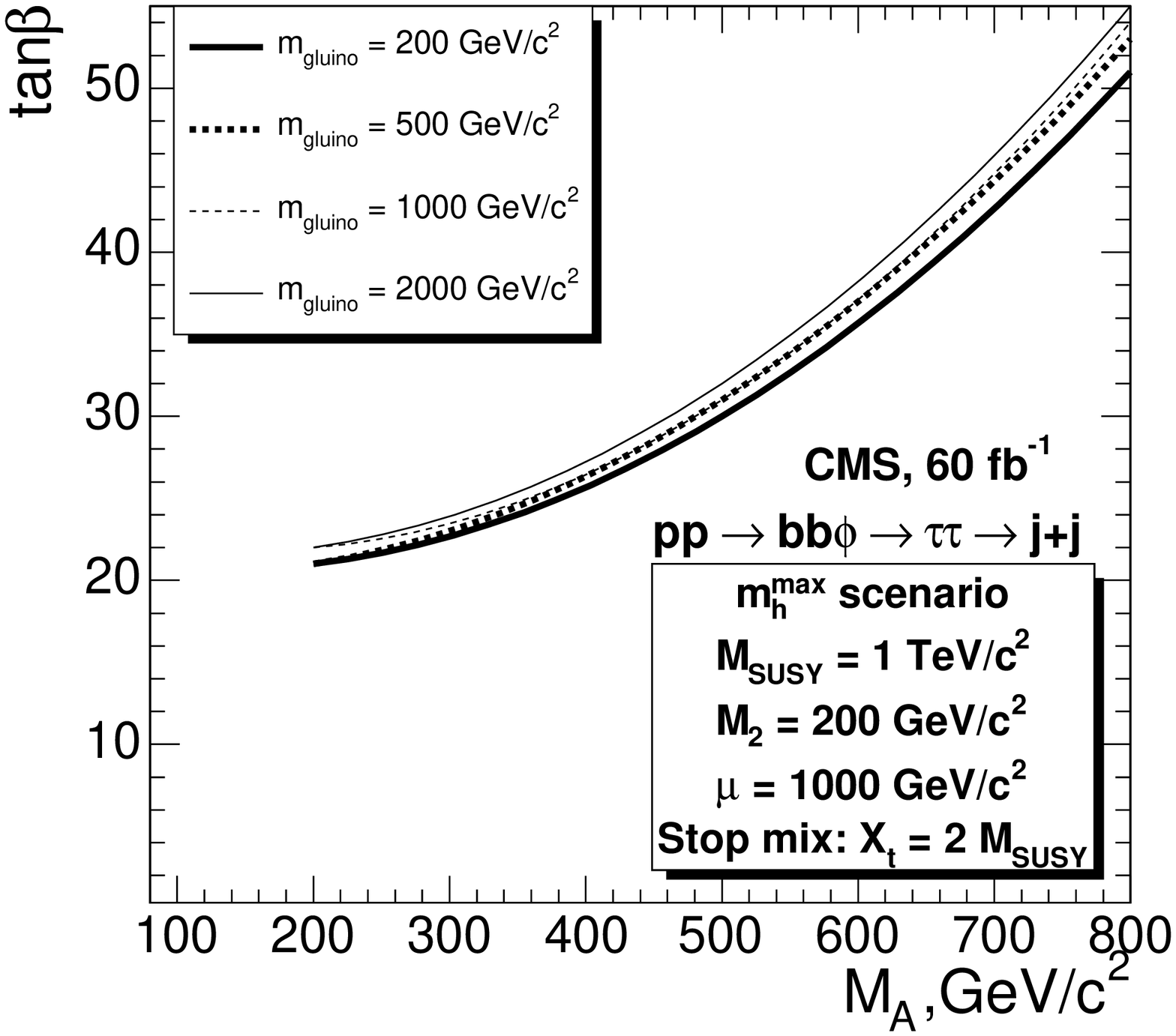}
\includegraphics[width=.49\textwidth]{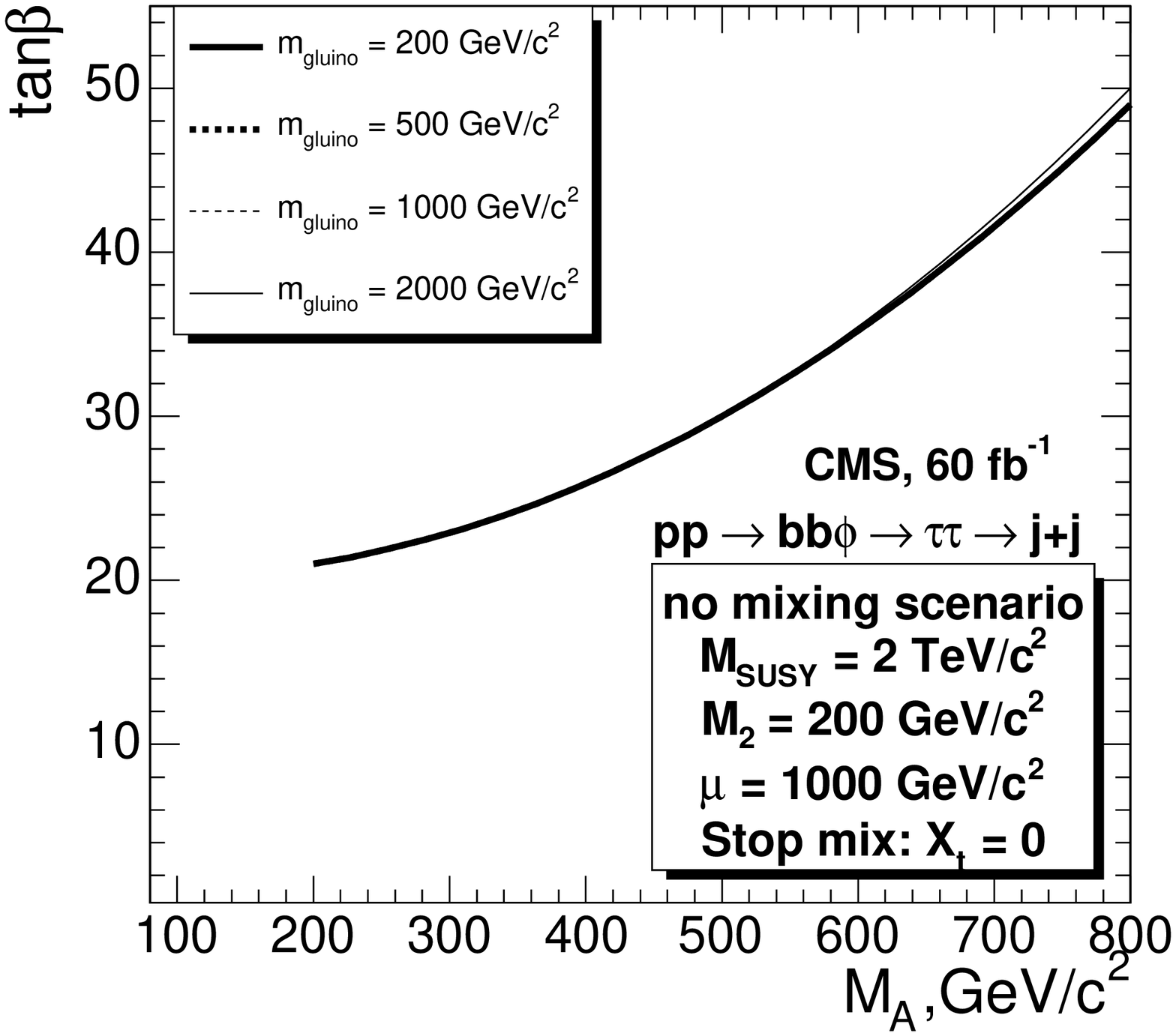}
\EC
\caption{Variation of the $5 \si$ discovery contours obtained 
from the channel $b\bar b \phi, \phi \to \tau^+\tau^- \to \,\mbox{jets}$
in the $\mhmax$ (left) and no-mixing (right) benchmark scenarios 
with $\mu = +1000 \gev$
for different values of~$\mgl$.}
\label{fig:jj_mu1000}
\end{figure}

A parameter affecting the $\db$ corrections, see \refeq{def:dmb}, but
not the kinematics of the Higgs-boson decays is the gluino mass, $\mgl$.
We now investigate the impact of varying this parameter, which is normally 
fixed to the values $\mgl = 800, 1600 \gev$ in the $\mhmax$ and no-mixing
benchmark scenarios, respectively. 
The results for four different values of the gluino mass, 
$\mgl = 200, 500, 1000, 2000 \gev$, are shown in \reffi{fig:jj_mu1000}.
The $\mu$ parameter has been set to $\mu = +1000 \gev$ in
\reffi{fig:jj_mu1000}, such that the Higgs decay channels into charginos
and neutralinos are suppressed. 
As one can see from \refeq{def:dmb}, the change of $\mgl$ affects the 
\order{\als} part of $\db$ and corresponds
to a monotonous increase of $\db$. As an example, this yields for
$\mu = 1000 \gev$, $\tb = 50$ in the two scenarios:
\BEA
\mhmax, \quad \mgl = 200 \gev &:& \db = 0.50 \non \\
\mhmax, \quad \mgl = 2000 \gev &:& \db = 0.94 \non \\
\mbox{no-mixing}, \quad \mgl = 200 \gev &:& \db = 0.06 \non \\
\mbox{no-mixing}, \quad \mgl = 2000 \gev &:& \db = 0.29~.
\EEA
In the no-mixing scenario the $\At$ value is close to zero, suppressing
the $\mgl$-independent contribution to $\db$, while the higher SUSY mass
scale results in an overall reduction of $\db$ in this scenario.
The value of $\db$ in the no-mixing scenario would slightly increase 
if $\mgl$ were raised to even larger values, but this effect would not
change the qualitative behaviour.

\reffi{fig:jj_mu1000} shows that the results
for the discovery reach in the $\MA$--$\tb$ plane are relatively stable 
with respect to variations of the gluino mass. The shift in the discovery 
contours remains below about $\De\tb = 4$ for the $\mhmax$ scenario
(left plot) and $\De\tb = 1$ for the no-mixing scenario (right plot).
For the positive sign of $\mu$ chosen in \reffi{fig:jj_mu1000}, where
the $\db$ correction yields a suppression of the bottom Yukawa coupling,
the largest discovery reach is obtained for small $\mgl$, while the
smallest discovery reach is obtained for large $\mgl$. This behaviour
would be reversed by a change of sign of $\mu$.

We have also investigated the possible impact of other MSSM parameters
(besides $\mu$ and $\mgl$) on the 5$\,\si$ discovery contours in the
$\MA$--$\tb$ plane. The $\db$ corrections depend also on the parameters
in the stop and sbottom sector, see \refeq{def:dmb}. While the formulas
in \refse{sec:db} have been given for the region where 
$\msusy \gg \mt$, the qualitative effect of reducing the stop and
sbottom masses can nevertheless be inferred. Sizable $\db$ corrections 
require relative large values of $\mu$ and $\mgl$. If these parameters
are kept large while the stop and sbottom masses are reduced, the $\db$
corrections tend to decrease. It is obvious from \refeq{def:dmb} that 
reducing the absolute value of $\At$ decreases the electroweak part of
the $\db$ correction. 
As discussed above, this effect of the $\db$ corrections manifests
itself in the
comparison of the $\mhmax$ and no-mixing scenarios, see
\reffis{fig:jj}--\ref{fig:jj_mu1000}.
Concerning the possible impact of the $\db$ corrections on the 5$\,\si$
discovery contours for the 
$b\bar b \phi, \phi \to \tau^+\tau^-$
channel in the $\MA$--$\tb$ plane we conclude that larger effects 
than those shown in 
\reffis{fig:jj}--\ref{fig:jj_mu1000} (where we have displayed the
discovery contours up to $\tb = 50$) would only arise if the variation
of $\mu$ were extended over an even wider interval than 
$-1000 \gev \leq \mu \leq +1000 \gev$ as done in our analysis above.

We now turn to the possible effects of other higher-order corrections
beyond those entering via $\db$ on the 5$\,\si$ discovery contours for 
the $b\bar b \phi, \phi \to \tau^+\tau^-$ channel.
These effects are in general non-negligible, see the discussions in 
\refse{sec:higherorder} and in \refse{sec:MAprec} below, but smaller 
than those induced by $\db$. As a consequence, the impact 
on the 5$\,\si$ discovery contours in the $\MA$--$\tb$ plane
of other
supersymmetric parameters entering via higher-order corrections 
is in general much smaller than the effect of varying $\mu$ in the 
high-$\tb$ region of \reffi{fig:jj_brchi0}. As an example, the
difference observed in \reffis{fig:jj}--\ref{fig:jj_mu1000} 
between the $\mhmax$ and no-mixing scenarios arising
from the different values of $\At$ and $\msusy$ in the two scenarios
(see \refeqs{mhmax}, (\ref{nomix})) is mainly an effect of the $\db$
corrections, while the impact of other higher-order corrections
involving $\At$ and $\msusy$ is found to be small.

Also the decays of the heavy neutral MSSM Higgs bosons into
supersymmetric particles are in general affected by other supersymmetric
parameters in addition to the dependence on $\mu$, $\MA$ and $\tb$.
The resulting effects on $\br(\phi \to \tau^+\tau^-)$ turn out to be
rather small, however. We find that sizable deviations from the values
of $\br(\phi \to \tau^+\tau^-)$ occurring in the $\mhmax$ and no-mixing
scenarios for $-1000 \gev \leq \mu \leq +1000 \gev$ are only possible in
quite extreme regions of the MSSM parameter space that are already
highly constrained by existing experimental data.

Our discussion above has been given in the context of the MSSM with real
parameters. Since the sensitivity of the 5$\,\si$ discovery contours in 
the $\MA$--$\tb$ plane on the other supersymmetric parameters can 
mainly be understood as an effect of higher-order corrections to the
bottom Yukawa coupling and of the kinematics of Higgs-boson decays into
supersymmetric particles, no qualitative changes of our results are
expected for the case where complex phases are taken into account.


\subsection{Higgs-boson mass precision}
\label{sec:MAprec}

The discussion in the previous section shows that the prospective
discovery reach of the 
$b\bar b \phi, \phi \to \tau^+\tau^-$ channel
in the $\MA$--$\tb$ plane is rather stable with respect to variations of the
other MSSM parameters. We now turn to the second part of our analysis
and investigate the expected statistical precision of the Higgs-boson
mass measurement. The expected statistical precision is evaluated as 
described in \refse{sec:expanal}, see \refeq{eq:precM}. 
In \reffis{fig:jjmass} -- \ref{fig:ejmass} we show the expected
precision for the mass measurement 
achievable from the channel 
$b\bar b \phi, \phi \to \tau^+\tau^-$ using the final states
$\tau^+\tau^- \to \,\mbox{jets}$ and  
$\tau^+\tau^- \to e + \,\mbox{jet}$. Within the 5$\,\si$ discovery 
region we have indicated contour lines corresponding to different values
of the expected precision, $\De M/M$.
The results are shown in the $\mhmax$ benchmark
scenario for $\mu = -200\gev$ (left plots) and $\mu = +200 \gev$ (right
plots).

We find that
experimental precisions of $\De M_\phi/M_\phi$ of 1--4\% are
reachable within the discovery region.
A better precision is reached for larger $\tb$ and smaller
$\MA$ as a consequence of the higher number of signal events in this
region. 
The other scenarios and
other values of $\mu$ discussed above 
yield qualitatively similar results to those
shown in \reffis{fig:jjmass}, \ref{fig:ejmass}.

\begin{figure}[htb!]
\vspace{1em}
\BC
\includegraphics[width=.49\textwidth]{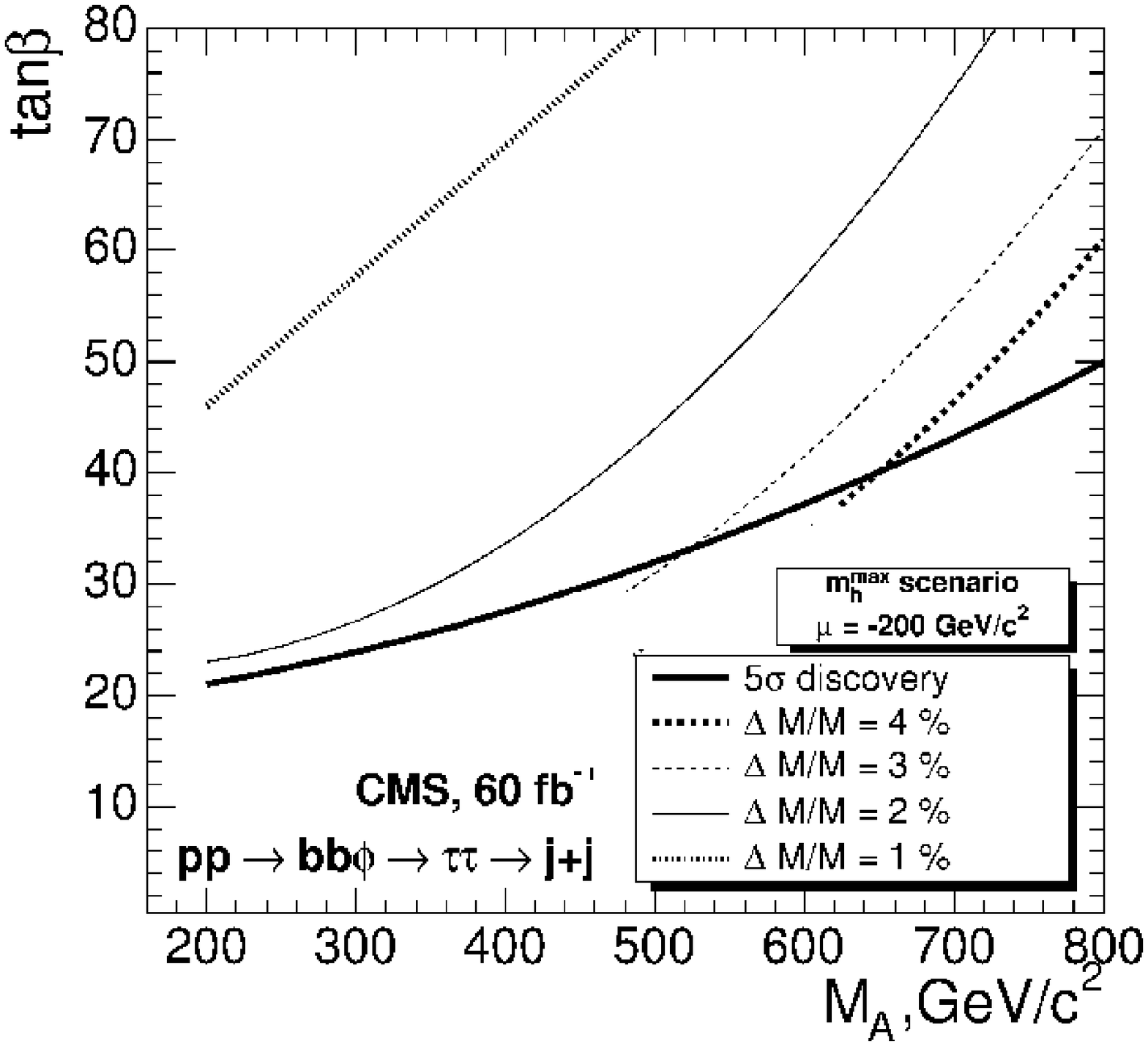}
\includegraphics[width=.49\textwidth]{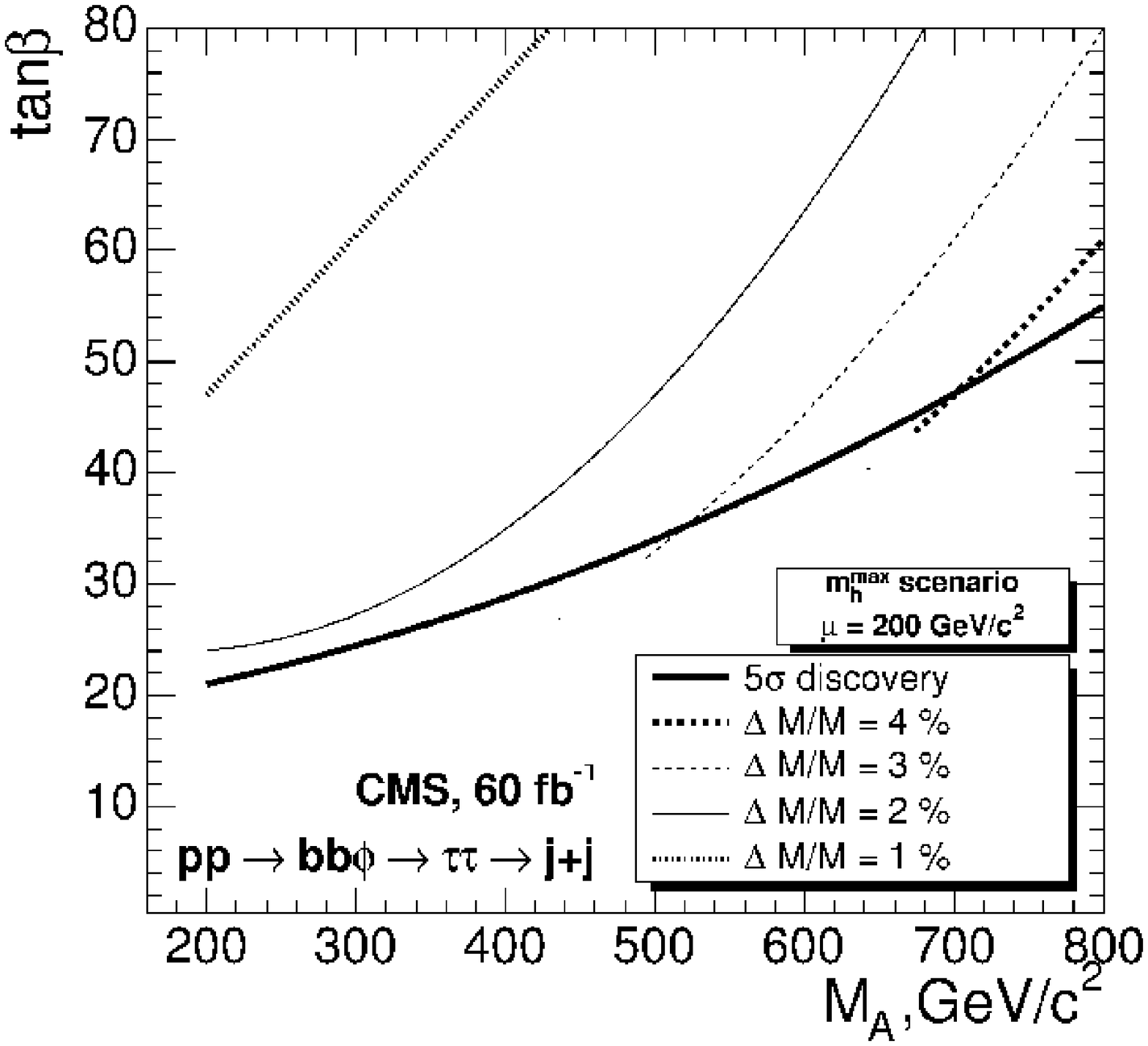}
\EC
\caption{The statistical precision of the Higgs-boson mass measurement
achievable 
from the channel $b\bar b \phi, \phi \to \tau^+\tau^- \to \,\mbox{jets}$
in the $\mhmax$ benchmark scenario for $\mu = -200 \gev$ (left) and 
$\mu = +200 \gev$ (right) is shown together with the 5$\,\si$ discovery 
contour.}
\label{fig:jjmass}
\vspace{1em}
\end{figure}


\begin{figure}[htb!]
\vspace{1em}
\BC
\includegraphics[width=.49\textwidth]{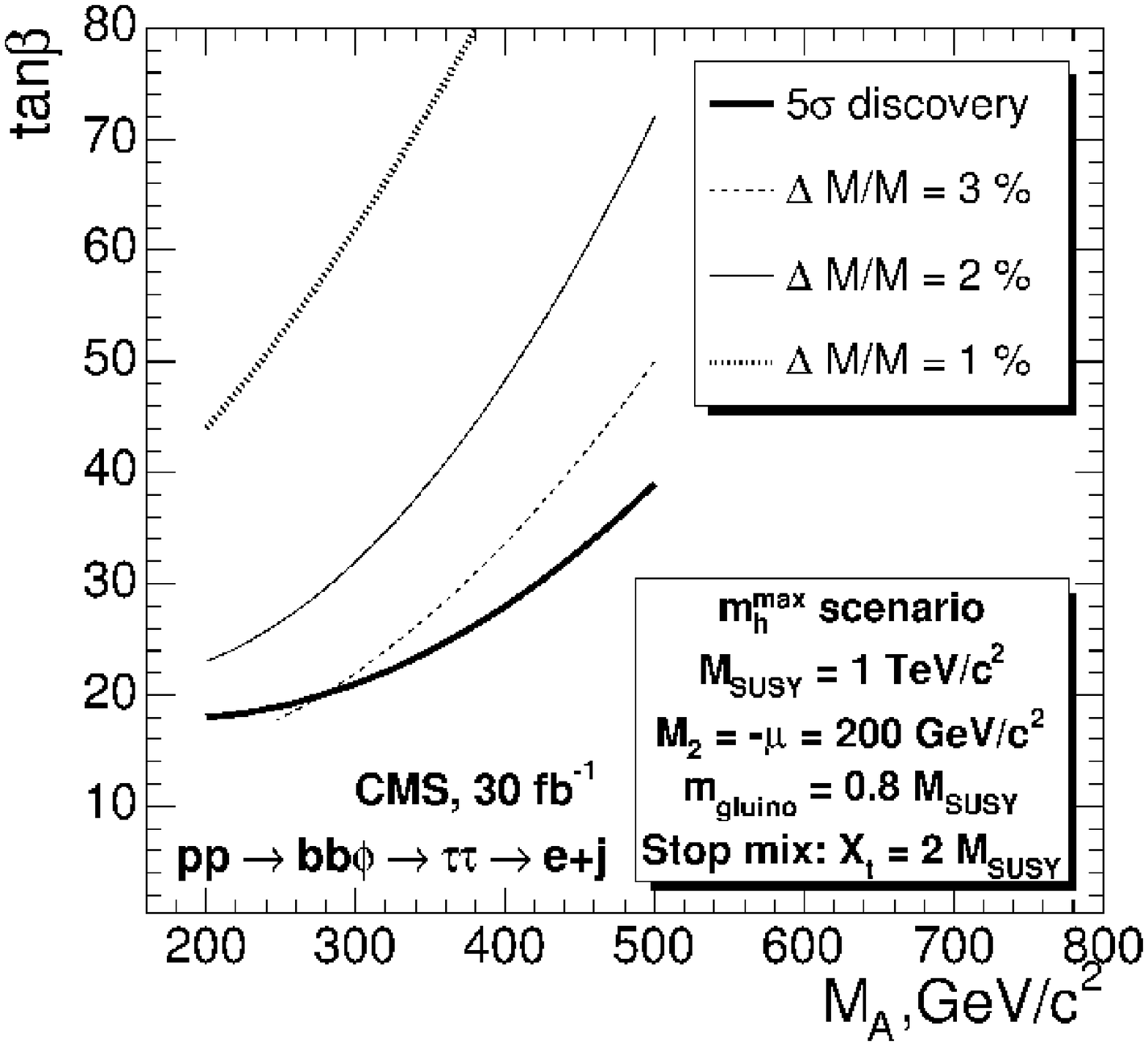}
\includegraphics[width=.49\textwidth]{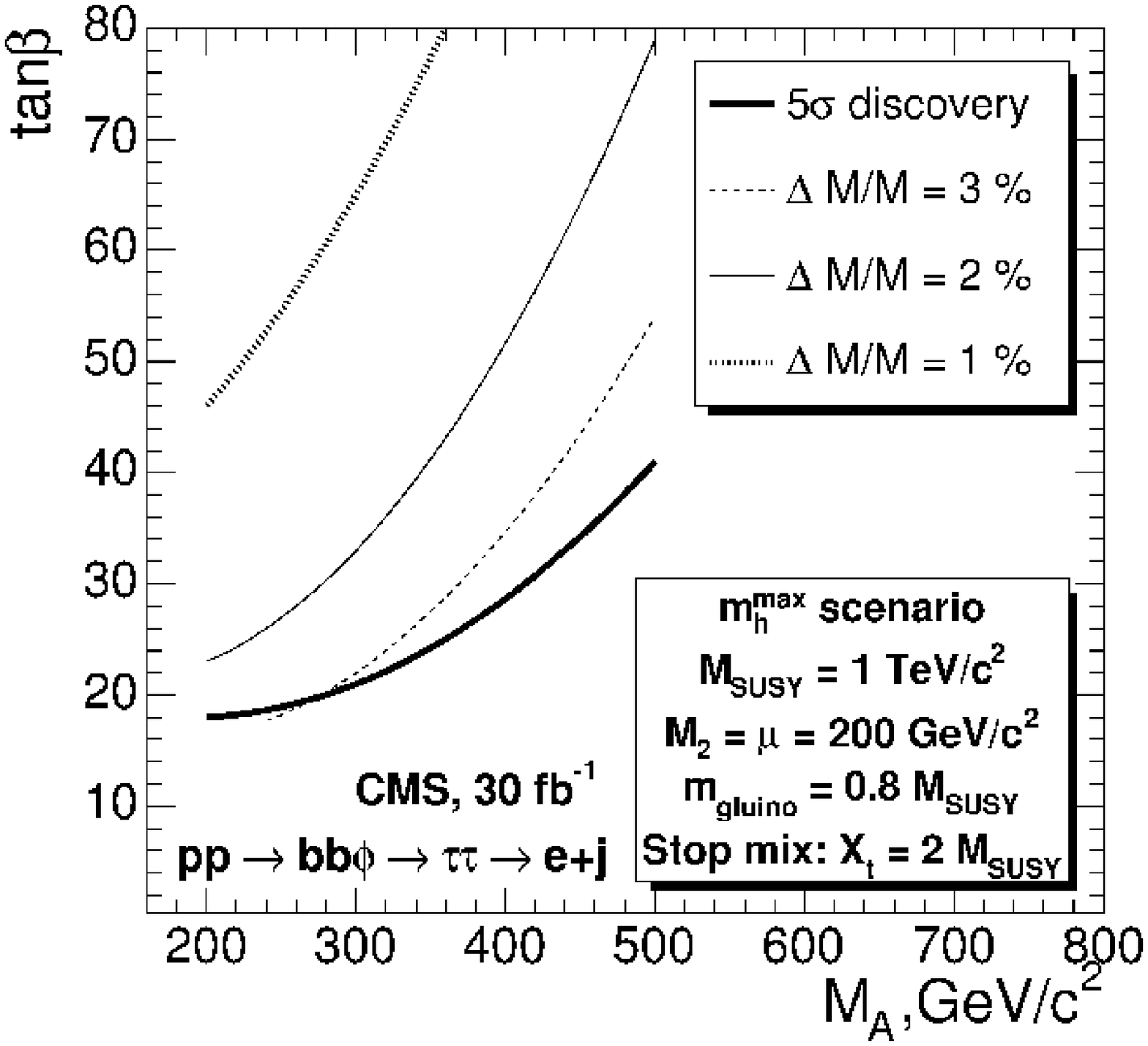}
\EC
\caption{The statistical precision of the Higgs-boson mass measurement
achievable
from the channel $b\bar b \phi, \phi \to \tau^+\tau^- \to e + \,\mbox{jet}$
in the $\mhmax$ benchmark scenario for $\mu = -200 \gev$ (left) and 
$\mu = +200 \gev$ (right) is shown together with the 5$\,\si$ discovery 
contour.}
\label{fig:ejmass}
\vspace{1em}
\end{figure}

\begin{figure}[htb!]
\vspace{1em}
\BC
\includegraphics[width=.65\textwidth]{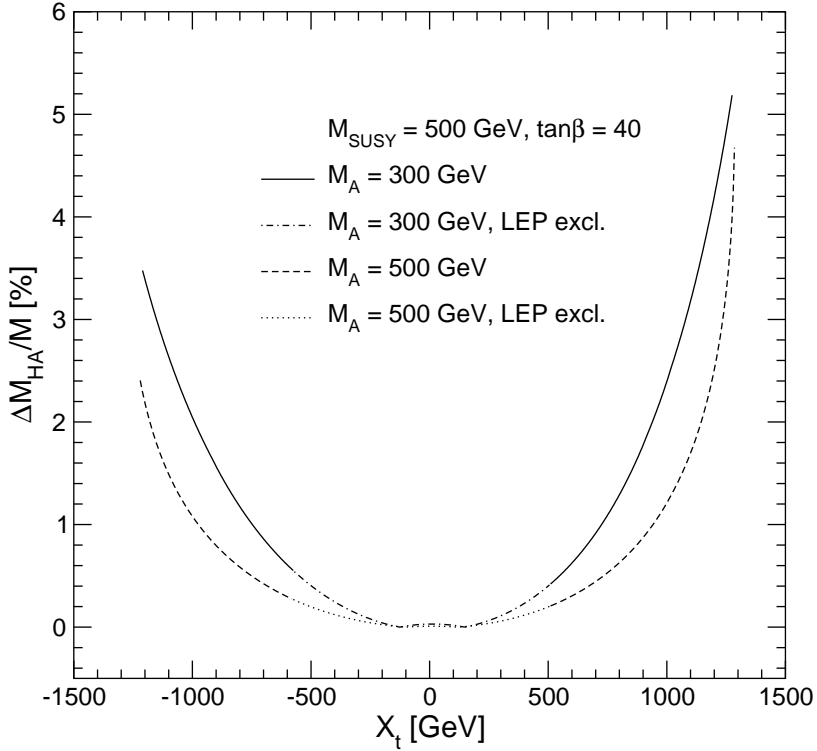}
\EC
\caption{The mass splitting between the heavy neutral MSSM Higgs bosons,
$\De M_{HA}/M \equiv \left|\MH - \MA\right|/\mbox{min}(\MH, \MA)$, is 
shown as a function of $\Xt$ for $\MA = 300, 500 \gev$ in a scenario with 
$\msusy = 500 \gev$, $\mu = 1000 \gev$ and $\tb = 40$. The 
other parameters are given in \refeq{eq:MAMHsplit}.
The dot-dashed (dotted) parts of the contours for $\MA = 300 \gev$
($\MA = 500 \gev$) 
indicate parameter combinations that are excluded by the search for the
light $\cp$-even Higgs boson of the MSSM at LEP~\cite{LEPHiggsMSSM}.}
\label{fig:massprec}
\vspace{1em}
\end{figure}

As discussed above, for large values of $\MA$ the heavy neutral MSSM
Higgs bosons are nearly mass-degenerate, $\MH \approx \MA$. The
experimental separation of the two states $H$ and $A$ (or the
corresponding mass eigenstates in the $\cp$-violating case) will
therefore be challenging. The results shown in 
\reffis{fig:jjmass} -- \ref{fig:ejmass} have been obtained using the
combined sample of $H$ and $A$ events. It is important to note, however, 
that even in the region of large $\MA$ the mass splitting between $\MH$
and $\MH$ can reach the level of a few \%. An example of such a scenario
is (as above, we consider the $\cp$-conserving case, i.e.\ the MSSM with
real parameters; the corresponding scenario in the case of non-vanishing
complex phases has been discussed in \citere{feynhiggs2.5})
\BEA
&&
\msusy = 500 \gev, \quad \At = \Ab = 1000 \gev, \quad \mu = 1000 \gev,  \non \\
&& \quad M_2 = 500 \gev, \quad M_1 = 250 \gev,
\quad \mgl = 500 \gev ~.
\label{eq:MAMHsplit}
\EEA
In \reffi{fig:massprec} the mass splitting 
\begin{equation}
\frac{\De M_{HA}}{M} \equiv \frac{\left|\MH - \MA\right|}{\mbox{min}(\MH, \MA)}
\end{equation}
is given as a function of $\Xt$ for $\tb = 40$ and two $\MA$ values,
$\MA = 300 \gev$ (solid line) and $\MA = 500 \gev$ (dashed line).
The dot-dashed and dotted parts of the contours for 
$\MA = 300, 500 \gev$, respectively, in the region of small $|\Xt|$
indicate parameter combinations that result in relatively low $\Mh$ 
values that are excluded by the search for the 
light $\cp$-even Higgs boson of the MSSM at LEP~\cite{LEPHiggsMSSM}.
One can see in \reffi{fig:massprec} that the mass splitting between
$\MH$ and $\MA$ shows a pronounced dependence on $\Xt$ in this scenario. 
Mass differences of up to 5\% are possible for large $\Xt$
(while the widths of the Higgs bosons are at the 1--1.5\% level in this
parameter region).

The example of \reffi{fig:massprec} shows that a precise mass
measurement at the LHC may in favourable regions of the MSSM parameter
space open the exciting possibility to distinguish between the signals
of $H$ and $A$ production. In confronting \reffi{fig:massprec} with the
expected accuracies obtained in \reffis{fig:jjmass} -- \ref{fig:ejmass} 
one of course needs to take into account that a separate treatment of
the $H$ and $A$ channels in \reffis{fig:jjmass} -- \ref{fig:ejmass}
would reduce the number of signal events by a factor of 2, resulting in
a degradation of the expected accuracies (for the same luminosity) by a
factor of $\sqrt{2}$. A more detailed analysis of the potential for
experimentally resolving two mass peaks would furthermore have to 
include effects arising from overlapping Higgs signals. Such an analysis
goes beyond the scope of the present paper.


\section{Conclusions}
\label{sec:conclusions}

We have analyzed the reach of the CMS experiment with 30 or 60~\ifb\ for the
heavy neutral MSSM Higgs bosons, depending on $\tb$ and the Higgs-boson mass
scale, $\MA$. We have focused on the channel
$b\bar b H/A, H/A \to \tau^+\tau^-$
with the $\tau$'s subsequently
decaying to jets and/or leptons. 
The experimental analysis, yielding the number of events
needed for a 5$\,\si$ discovery (depending on the mass of the Higgs
boson) was performed with full CMS detector 
simulation and reconstruction for the final states of di-$\tau$-lepton decays.
The events were generated with PYTHIA. 

The experimental analysis has been combined with predictions for the
Higgs-boson masses, production processes and decay channels obtained
with the code {\tt FeynHiggs}, taking into account all relevant
higher-order corrections as well as possible decays of the heavy Higgs 
bosons into supersymmetric particles. We have analyzed the sensitivity
of the 5$\,\si$ discovery contours in the $\MA$--$\tb$ plane to
variations of the other supersymmetric parameters. We have shown that
the discovery contours are relatively stable with respect to the impact of 
additional
parameters. The biggest effects, resulting from higher-order
corrections to the bottom Yukawa coupling and from the kinematics of
Higgs decays into charginos and neutralinos, are caused by varying the 
absolute value and the sign of the
higgsino mass parameter $\mu$. The corresponding shift in the 5$\,\si$
discovery contours amounts up to about $\De\tb = 10$. The effects of
other contributions to the relation between the bottom-quark mass 
and the bottom Yukawa coupling, arising from the gluino mass and the 
parameters in the stop and sbottom sector, are in general smaller than
the shifts induced by a variation of $\mu$. The same holds for the
impact of higher-order contributions beyond the corrections to the
bottom Yukawa coupling and for the possible effects of other decay modes
of the heavy Higgs bosons into supersymmetric particles. The results of
our analysis, which was carried out in the framework of the
$\cp$-conserving MSSM, should not be substantially affected by
the inclusion of complex phases of the soft-breaking parameters.

We have analyzed the prospective accuracy of the mass measurement of the
heavy neutral MSSM Higgs bosons in the channel
$b\bar b H/A, H/A \to \tau^+\tau^-$. We find that statistical
experimental precisions of 1--4\% are
reachable within the discovery region. These results, obtained from a
simple estimate of the prospective accuracies, are not expected to 
considerably degrade if further uncertainties related to background
effects and jet and missing $E_{\rm T}$ scales are taken into account.
We have pointed out that a \%-level precision of the mass measurements 
could in favourable regions of the MSSM parameter allow to 
experimentally resolve the signals of the two heavy MSSM Higgs bosons.


\subsection*{Acknowledgements}

S.H.\ and G.W.\ thank M.~Carena and C.E.M.~Wagner for collaboration on some of
the theoretical aspects employed in this analysis.


\newpage
\bibliographystyle{plain}


\end{document}